# Efficient Inverse Design of Plasmonic Patch Nanoantennas using Deep Learning

**Saeed Hemayat,**[a†] **Sina Moayed Baharlou,**[a,b†] **Alexander Sergienko,**[b] **and Abdoulaye Ndao**[a,b*]

[a] Department of Electrical and Computer Engineering, University of California San Diego, La Jolla, CA 92093, USA
[b] Department of Electrical and Computer Engineering and Photonics Center, Boston University, 8 Saint Mary's Street, Boston, MA 02215, USA
* Corresponding author, Email: *a1ndao@ucsd.edu*
[†] Equal contribution

**Abstract.** Plasmonic nanoantennas with suitable far-field characteristics are of huge interest for utilization in optical wireless links, inter-/intra-chip communications, LiDARs, and photonic integrated circuits due to their exceptional modal confinement. Despite its success in shaping robust antenna design theories in radio frequency and millimeter-wave regimes, conventional transmission line theory finds its validity diminished in the optical frequencies, leading to a noticeable void in a generalized theory for antenna design in the optical domain. By utilizing neural networks, and through a one-time training of the network, one can transform the plasmonic nanoantennas design into an automated, data-driven task. In this work, we have developed a multi-head deep convolutional neural network serving as an efficient inverse-design framework for plasmonic patch nanoantennas. Our framework is designed with the main goal of determining the optimal geometries of nanoantennas to achieve the desired (inquired by the designer) $S_{11}$ and radiation pattern simultaneously. The proposed approach preserves the one-to-many mappings, enabling us to generate diverse designs. In addition, apart from the primary fabrication limitations that were considered while generating the dataset, further design and fabrication constraints can also be applied after the training process. In addition to possessing an exceptionally rapid surrogate solver capable of predicting $S_{11}$ and radiation patterns throughout the entire design frequency spectrum, we are introducing what we believe to be the pioneering inverse design network. This network enables the creation of efficient plasmonic antennas while concurrently accommodating customizable queries for both $S_{11}$ and radiation patterns, achieving remarkable accuracy within a single network framework. Our framework is capable of designing a wide range of devices, including single band, dual band (two uncorrelated frequencies), and broadband antennas, with directivities and radiation efficiencies reaching 11.07 dBi and 75%, respectively, for a single patch. The proposed approach has been developed as a transformative shift in the inverse design of photonics components, with its impact extending beyond antenna design, opening a new paradigm toward real-time design of application-specific nanophotonic devices.

## 1 Introduction

Control and manipulation of light at the nanoscale is considered as one of the cornerstones of modern optics, with the potential to revolutionize scientific and technological advances. Through the years, light manipulation has been implemented through various approaches such as photonic crystals [1], [2], metamaterials [3], [4], metasurfaces [5], [6], [7], [8], [9], [10], [11], [12], [13], [14], [15] and plasmonic structures due to their unprecedented ability to locally control and manipulate the incident light at the nanoscale [16], [17], [18], [19], [20]. Plasmonic nanoantennas serve a crucial role in a wide range of applications such as plasmonic lenses [21], [22], plasmonic tweezers [23], [24], [25], intra/inter-chip optical communications [26], [27], LiDARs [28], augmented reality and holography [29], [30], imaging [31], and in surface-enhanced Raman spectroscopy (SERS) [32] due to their unique ability to guide and confine light at the nanoscale. To date, plasmonic nanoantennas are mainly used for near-field applications and lack a robust far-field performance and a generalized far-field design methodology. Although conventional antenna theory has been successful in shaping the design theory and techniques in low-frequency regimes such as radio frequency (RF) or mm-wave, as we venture towards the optical domain, due to the radically different wave-matter interactions, the validity of this theory diminishes significantly [33]. By leveraging neural networks, we can convert this problem into an automated, data-driven task.

Data-driven methods such as deep neural networks (DNNs) are receiving significant attention owing to their remarkable success in computer vision [34], [35], natural language processing [36], [37], and speech recognition [38]. In nanophotonics, DNNs have been used to replace the complex and time-consuming design procedures by approximating the electromagnetic simulations and learning the inverse process [39], [40], [41], [42], [43], [44], [45], [46], [47], [48], [49], [50], [51], predicting the fabrication imperfections [52], and post-fabrication appearance [53]. While promising, DNNs face challenges with inverse problems due to their reliance on a large number of labeled samples (i.e., devices with simulated responses), which grow exponentially with additional degrees of freedom of the device. Also, discriminative neural networks may lead to sub-optimal results due to the existing non-uniqueness in inverse problems. Prior studies addressed the inverse design problem using discriminative networks in combination with brute forcing [54], analytical gradient [41], and evolutionary algorithms [55], [56], [57]. Tandem networks have been utilized in various works [58], [59], and generative models such as variational autoencoders [60], [61] and generative adversarial networks [43], [62], [63], [64] have been adapted to enhance the design with more degrees of freedom. However, these methods face severe constraints, such as inadvertent discarding of desirable devices in tandem models due to transforming the one-to-many mappings to one-to-one mappings, challenges in encapsulating fabrication constraints, and difficulties in training generative models that may lead to blurry and inaccurate results [65]. In addition, the generative models suffer from mode collapses, limiting their ability to generate multiple diverse results.

In this work, we have developed an inverse-design framework for efficiently designing plasmonic patch nanoantennas to overcome the aforementioned obstacles. Our framework is capable of determining the optimal configuration of nanoantennas to achieve the desired and physically possible $S_{11}$ and radiation pattern. The proposed framework is developed based on the pseudo-inverse function. It utilizes a multi-head deep convolutional neural network as a surrogate solver to accurately estimate the $S_{11}$ and the radiation pattern of a given device across the entire frequency range. This is orders of magnitude faster than numerical simulations. The particle swarm optimization (PSO) is used in conjunction with the surrogate solver to efficiently search the design space and locate the desired devices. Following the search, a clustering algorithm is applied to identify multiple diverse results. Contrary to most NN-based inverse-design methods, our proposed approach preserves the one-to-many mappings. It allows the designer to choose from multiple diverse devices for a given design problem. The framework enables the designer to add fabrication constraints even after the training process and generate the desired devices through complex queries, enhancing customization in the design process. To the best of our knowledge, this is the first time that a neural network-based inverse design framework encapsulates all of the mentioned properties while maintaining simplicity and fast runtimes. The proposed framework can design a wide range of devices with the desired characteristics, including single band, dual band, and broadband antennas with a maximum directivity of up to 11.07 dBi and radiation efficiencies reaching almost 75% for a single patch. The proposed approach has been developed to serve as a transformative foundation in inverse design, with its impact extending beyond antenna design and toward real-time design of application-specific nanophotonic devices.

## 2 Deep Learning-Based Inverse Design Framework

The conventional design process starts with a structure or a set of known input parameters and obtains the corresponding outcome afterward. However, in the inverse design, the process works the other way around. The designer starts with a set of known desired outputs, and the goal is to discover the structure or parameters that can produce those specific outcomes. This process is of great importance because automating it with artificial intelligence (AI) and data-driven methods can significantly accelerate the design process and save time and resources. Additionally, it is possible to identify new and previously unattainable devices that outperform the existing solutions.

Our inverse design framework utilizes a deep neural network (serving as a surrogate solver) to model the simulation process and uses PSO to search the design space. The surrogate solver replaces the computationally intensive numerical simulation process, enabling the PSO algorithm to explore the design space efficiently and identify the devices with desired responses. The proposed framework comprises three components: the "Multi-head Convolutional Surrogate Solver," the "Particle Swarm Optimization Algorithm," and the "Clustering Algorithm" (as shown in Figure 1(b)). Together, these components generate a collection of feasible and manufacturable nanoantennas with desired responses. The framework takes a desired response in a form of a query (Figure 1(a)) and generates a set of devices that exhibit those responses (Figure 1(c)). The query consists of a set of high-level conditions the desired device must meet. Each condition is represented as a cost function that the PSO aims to minimize (named design objectives). In addition to the query, the designer can define fabrication constraints and clustering parameters to specify the extraction of multiple devices. Fabrication constraints can be incorporated into the inverse process by either adding them as an extra cost function to the PSO objective function or by limiting the search space of the PSO algorithm. After defining the search space and the objective function, PSO begins optimizing the objective function: in other words, PSO will search the device space for a set of devices that meets the requirements. In the search process, several devices need to be evaluated (i.e., their responses should be simulated). This evaluation is done using the multi-head deep convolutional surrogate solver, which is trained to approximate the responses of the given device. The architectural detail of the surrogate solver is illustrated in Figure 1(d). The surrogate solver enables the PSO to search the device space in a matter of seconds. PSO iteratively collaborates with the surrogate solver until convergence. After the device space is searched, the mean shift algorithm clusters the resulting particles to locate a set of admissible solutions, instead of just one. It then outputs the set of devices, as illustrated in Figure 1(c). The surrogate solver and the inverse process are explained in detail in sections 2.3 and 2.4, respectively.

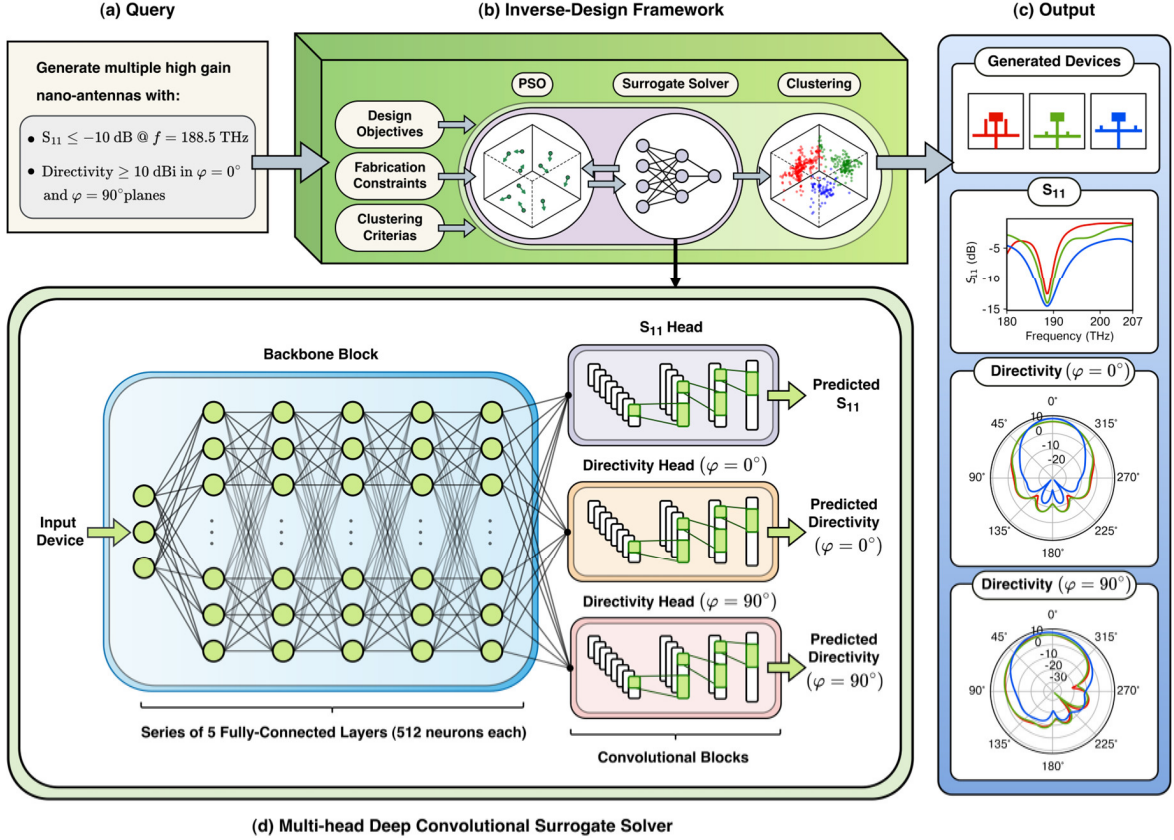

**Fig. 1.** An overview of the proposed inverse-design framework. Our approach is based on solving the pseudo-inverse function. It employs a deep neural network as a surrogate solver to replace the computationally expensive and time-consuming simulation process and uses an optimization algorithm to search the design space for the desired devices. Our approach preserves the one-to-many mappings, allowing the generation of multiple devices and accommodating fabrication constraints. The proposed framework operates by taking a requested response in the form of a query and a set of fabrication constraints as input. An objective function is then defined based on these inputs, which must be minimized using the PSO. A multi-head deep convolutional neural network is trained to perform as a surrogate solver and quickly predict the device responses, including the device $S_{11}$ and radiation patterns. The PSO efficiently explores the design space using the surrogate solver and identifies the desired devices. Finally, a clustering algorithm is applied to generate a diverse set of multiple devices. (a) The desired responses in the form of a query, (b) our inverse design framework, (c) generated devices given the desired response, (d) multi-head deep convolutional surrogate solver.

We refer to our approach as solving the pseudo-inverse function, which involves modeling the forward process using the neural network and optimizing it to find the desired devices. In the following, we will explain the pseudo-inverse function, its advantages, and why modeling the inverse process directly using neural networks has challenges and would not provide the mentioned benefits.

Assuming that the parameters and response of the nanoantenna can be represented as vectors $d \in \mathbb{R}^n$ and $r \in \mathbb{R}^m$, where $n$ and $m$ are the dimensions of the device and response space, respectively. We will define the function $f: d \to r$ as the forward design function. This function is known and well-defined, meaning that for any input device $d$, a single response $r$ is produced. This function can be evaluated through time-consuming numerical simulations.

Considering the above notations, the function $f^{-1}: r \to d$ is called the inverse design function. The ability to evaluate, learn, and estimate this function plays a crucial role in inverse design tasks because a target device with a set of desired responses can be determined by evaluating this function. However, $f$ is neither injective nor surjective, leading to the fact that it does not have a well-defined inverse function $f^{-1}$. As a result, some responses cannot be generated by any device, while multiple devices can generate others (see supplementary Figure S1), resulting in one-to-many mappings. When using machine learning models such as discriminative neural networks to model the inverse function, one of the main challenges is dealing with one-to-many mappings in the dataset. This is because discriminative neural networks are designed to learn one-to-one mappings, and modeling the inverse function using these networks would result in poor convergence and inaccurate results. A workaround to overcome this issue is converting $d \to r$ to a bijective mapping [58], resulting in a one-to-one inverse function that removes potentially valuable devices from the device space.

Contrary to methods that do not preserve the one-to-many mappings, our inverse design framework is based on solving the pseudo-inverse function, denoted as $f^\dagger$, with the property of $f(f^\dagger(r)) \approx r$, which can preserve the one-to-many mappings. This function is defined as $f^\dagger(r) = D$, where D is a set of possible solutions (devices), each satisfying the following condition:

$$D = \{d \in \mathbb{R}^n, \|f(d) - r\|^2 < \varepsilon\}, \tag{1}$$

where $\varepsilon$ is a predefined threshold value indicating the maximum allowed discrepancy between the desired and the target responses. A set of possible solutions $D$, can be obtained by a search or an optimization algorithm. This process includes multiple evaluations of the forward function (the numerical simulations), which can be time-consuming. To speed up this process, we have developed and trained a multi-head convolutional neural network as the surrogate solver to approximate the simulation function, allowing for a fast and parallel evaluation of $f$. Additionally, we have injected the designer's knowledge by incorporating a predefined structure and considering a set of prior fabrication constraints (such as minimum feature sizes and minimum distances between the T-stubs with the feed and the patch), reducing the device space significantly. As a result, our pseudo-inverse function can be defined as follows:

$$D = \left\{d \in \mathbb{R}^k, \|\hat{f}(d) - r\|^2 < \varepsilon\right\}, \tag{2}$$

where $k$ is the dimension of the reduced device space $k \ll n$, and $\hat{f}(d)$ is the approximated simulation function using the surrogate solver. The values of $d$ are also limited to the range of motion of the parameters of a predefined structure. Having a fast surrogate solver and a bounded device space, PSO is utilized consequently for efficient exploration of the device space and determining possible solutions $D$. Since the one-to-many mappings remain intact, multiple diverse solutions can be determined by identifying the clusters formed by the PSO algorithm using a clustering algorithm and obtaining the local minima in each cluster. Notably, the proposed surrogate solver performs the PSO's particle evaluation stages in parallel, allowing all particles to be evaluated simultaneously, which hugely increases the efficiency of the inverse algorithm. For instance, designing a single device using our framework requires an average of 20 iterations of the PSO, with 512 devices evaluated in each iteration. The average execution time of our inverse design process is $0.08 \pm 0.02$ seconds per device, depending on the hardware. It is noteworthy that, without the surrogate solver, the execution time to design one device would take more than 56 hours.

One of the most important advantages of the proposed approach lies in the fact that fabrication constraints can be imposed by either limiting the search space (e.g., fixing a parameter, reducing the range of a variable) or penalizing the regions where the fabrication constraints are not met, after the training and learning process (one does not need to limit the training dataset space to a dataset where fabrication constraints have already been applied, which will eliminate many of the potential devices). Furthermore, the designer can choose a less sensitive device to fabrication imperfections, as the one-to-many mapping is preserved and multiple solutions are discovered.

*2.1 Dataset*

Deep neural networks often require large-scale datasets for accurate predictions. The number of training samples varies based on different factors, including the input and output dimensions and the mapping complexity. Gathering a large-scale dataset is both time-consuming and costly, especially when training a physical surrogate solver, as the generated devices must also go through the simulation process.

In this work, by exploiting the designer's knowledge, we have significantly reduced the dimensions of the device space and have bound it to meet fabrication limitations, such as the overall size of the device, minimum feature sizes, minimum distance between two elements, and compatibility of the device with current fabrication technologies. Our devices have three or four degrees of freedom, and we are specifically interested in the device's $S_{11}$ response and radiation pattern. The $S_{11}$ frequency range spans from 180 THz to 207 THz, and we have sampled the data at 96 points within this range. Additionally, we have captured the directivity of the device at four frequencies of interest (185 THz, 188.5 THz, 193.5 THz, 198.5 THz). For each frequency, we have two cuts of the radiation pattern in $\varphi = 0°$ and $\varphi = 90°$, and this data has been sampled through 72 points. This makes our device space belong to $d \in \mathbb{R}^3$ or $d \in \mathbb{R}^4$, and our response space belongs to $r \in \mathbb{R}^{672}$. In the text, the responses are denoted individually in the form of $r_{S_{11}} \in \mathbb{R}^{96}, r_{\varphi=0°} \in \mathbb{R}^{4\times 72}, r_{\varphi=90°} \in \mathbb{R}^{4\times 72}$, or in a concatenated and flattened form of $r = \left[r_{S_{11}}^\top, r_{\varphi=0°}^\top, r_{\varphi=90°}^\top\right]^\top$.

A dataset of 50,000 samples has been generated and simulated with their corresponding responses for the device with three degrees of freedom using the CST Studio. Throughout this process, Latin hypercube sampling (LHS) [66] has been used to generate random samples due to its efficiency in covering the parameter space compared to simple random sampling (see Supplementary section S1.1 for the importance of using LHS). 90% of the generated data has been used for training (45,000 samples), and the remaining 10% has been kept for validation and testing (2,500 each). The validation set evaluates the surrogate model to obtain the optimal architecture and training hyperparameters. In contrast, the test set determines the model's final accuracy. We have also verified that all the generated samples are unique, with no leakage of validation or test sets.

Throughout the text, the experiments and the results are reported for the dataset with three degrees of freedom, and the quantitative and qualitative results for the device with four degrees of freedom can be found in the supplementary material.

Upon further analysis of the gathered dataset, the presence of one-to-many mappings was confirmed. This was achieved by extracting distinct devices from the dataset that exhibited similar responses (see supplementary Figure S1, which shows three instances of this relationship).

Furthermore, we have observed a strong linear correlation between the radiation patterns of different frequencies (on the same cut), indicating that the radiation pattern varies smoothly as the frequency changes (see supplementary Figure S2). We utilized this observation while designing our surrogate solver to determine the radiation pattern at other frequencies within the range of our interest through linear interpolation. More detailed description of this observation can be found in Supplementary section S1.4.

*2.2 Basic Antenna Design*

Infusion of the designer's knowledge into the inverse design process not only ensures that the design process is grounded in practicality and real-world applicability, but also significantly reduces the size of the required dataset. This is important particularly due to the complex nature of plasmonic systems, where generating large datasets is time-consuming and requires computationally expensive numerical simulations. Moreover, by integrating domain-specific knowledge, the network becomes more efficient at generalizing from smaller datasets. Here, to inject the designer's knowledge into the model, the basic structure of the antenna is designed using the well-known formulas for plasmonic patch antennas [26], [27], and two separate datasets with three and four parameters have been generated by adding T-stub configurations to the pre-designed basic structure.

In the optical regime, metals behave differently than in the radio frequency (RF) due to the negative values of the real part of their permittivity. This unique feature of metals enables them to support surface modes at the metal/insulator interface, namely the surface plasmon polaritons (SPPs). A plasmonic MIM waveguide is comprised of two metal/insulator interfaces where each metal/insulator interface supports individual SPPs. Bringing the two interfaces to the same proximity results in the coupling of the SPPs on the two interfaces and a single propagating plasmonic mode in the MIM waveguide. The propagating mode in a MIM plasmonic waveguide is a transverse magnetic (TM) in nature, and as a result it cannot be described or analyzed using the conventional transmission line theory (since the conventional transmission line theory considers all metals as perfect electric conductors (PECs), as a result boundary conditions do not allow any tangential electric fields, and consequently TM modes). However, one can use a mere approximative method (see supplementary section S2) and take advantage of the fact that the magnitude of the TM component is rather small with respect to the transverse component and approximate the mode as a TEM mode to design the basic parameters (it should be noted that the formulas and the methodologies outlined in supplementary section S2 are predominantly based on empirical and numerical fits that were previously proposed in the literature for plasmonic patch nanoantennas and these methodologies are applied here exclusively for the design of the basic structure of the antenna which we will use to generate the dataset, while our proposed inverse design framework will do further designs).

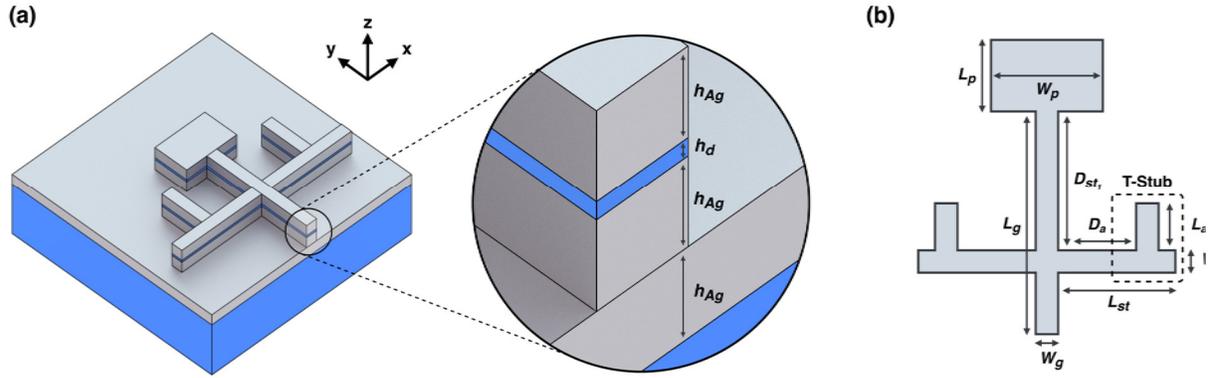

**Fig. 2.** Basic structure of the plasmonic patch. (a) 3D view of the full structure, (b) Top view of the antenna.

The basic parameters of the patch are chosen as shown in Figure 2, where $W_g$ = 100 nm, $L_g$ = 1000 nm, $L_{st}$ = 850 nm, $W_p$ = 500 nm, and $L_p$ = 320 nm. Our inverse network will determine $D_{st}$, $D_a$, and $L_a$. This selection aims to achieve impedance matching for single band, dual band, and broadband operations, using various shapes and configurations of the T-stub that will be further designed and added to the structure by our inverse design network. The thickness of the metallic layers in the waveguide and the patch have been chosen in a way to be larger than the surface wave skin depth $\delta_m$ (see Supplementary section S3), but not very small which leads to fabrication implications ($h_{Ag}$ = 100 nm), whereas the thickness of the dielectric layer is chosen as $h_d$ = 20 nm (see Supplementary section S4, Figure S3(b)).

Although the most prominent and most accessible method to control the resonant frequency of the patch is by controlling the length of the patch, a different approach has been chosen here. Here, the length of the patch is fixed, and we will tune the resonant frequency (resonant frequencies for dual band and broadband operations) of the patch using various symmetrical T-stub configurations. The basic shapes of the T-stubs are based on the T-shaped resonators that previously used in diplexers [67] and dual band transformers [68]. Here, we will show that, our network is capable of generating all of the desired responses (forward problem) and the devices (inverse problem) for all of the queries using two T-stubs (the three-parameter case), and a combination of two T-stubs with two normal stubs (stubs without additional arms) configurations (the four-parameter case), without changing the patch dimensions.

There are several reasons behind this choice: by altering the T-stub dimensions and locations, one can also control the antenna's bandwidth without significantly altering other antenna characteristics, whereas changing the patch size will not provide the same level of control over bandwidth. Additionally, although one of the widely accepted methods to induce dual band or broadband operation in the patches is introducing slots in the patch, this will increase the radiating edge of the patch and lead to higher edge currents, resulting in increased spurious radiation and decreased radiation efficiency. Moreover, slots might lead to higher cross-polarization levels; they require tighter tolerances during fabrication (especially in plasmonic structures where feature sizes are extremely small) as precise slot dimensions and positions are crucial for achieving the desired multi band performance, and slots are also typically hard to design as they can introduce additional resonances, which may lead to unwanted harmonic radiations. Apart from the drawbacks mentioned above, inducing multi band or broadband operation in the antenna typically using slots requires the creation of slots of different shapes and sizes in the patch and this method hugely increases the number of parameters in the hyperparameter space (e.g., length, width, shapes, and location of the slots), and the data required to train the network.

*2.3 Multi-head Convolutional Surrogate Solver*

As previously mentioned, the simulation of electromagnetic devices requires considerable computational resources and time. For instance, simulating a single nanoantenna can take over 20 seconds (with symmetric boundary conditions), even on a high-end computer. To facilitate this process, we have utilized deep neural networks to approximate the simulation function. Using deep neural networks as a surrogate solver exhibits several advantages: it significantly reduces computation time, performs tasks orders of magnitude faster, and enables parallel evaluation of multiple samples. Moreover, we can use backpropagation to compute the derivative of the simulation function, which is beneficial for optimization tasks.

The architecture of deep neural networks plays a vital role in their performance. Factors such as the number and type of layers and the activation function significantly impact the network's ability to generalize to unseen data. Additionally, selecting an appropriate inductive bias, such as convolution over fully connected, can reduce the required training samples. We have conducted an extensive hyper-parameter optimization (HPO) process to select the optimal network architecture and training parameters (e.g., learning rate, weight decay, batch size). This process involves sampling different configurations based on the pre-defined range of hyperparameters and a set of network configurations. Subsequently, the network is trained with the sampled configuration, its performance is evaluated on the validation set, and the best configuration is selected from the sampled configurations (each iteration of this process is called a trial). HPO can speed up the entire process with trial pruning and early stopping techniques, in which trials with less promising results are terminated earlier. The parameters considered in the HPO for sampling include the learning rate, regularization weight, batch size, configuration of fully connected layers (i.e., number of layers and neurons), whether to use convolutional layers, and their corresponding configurations.

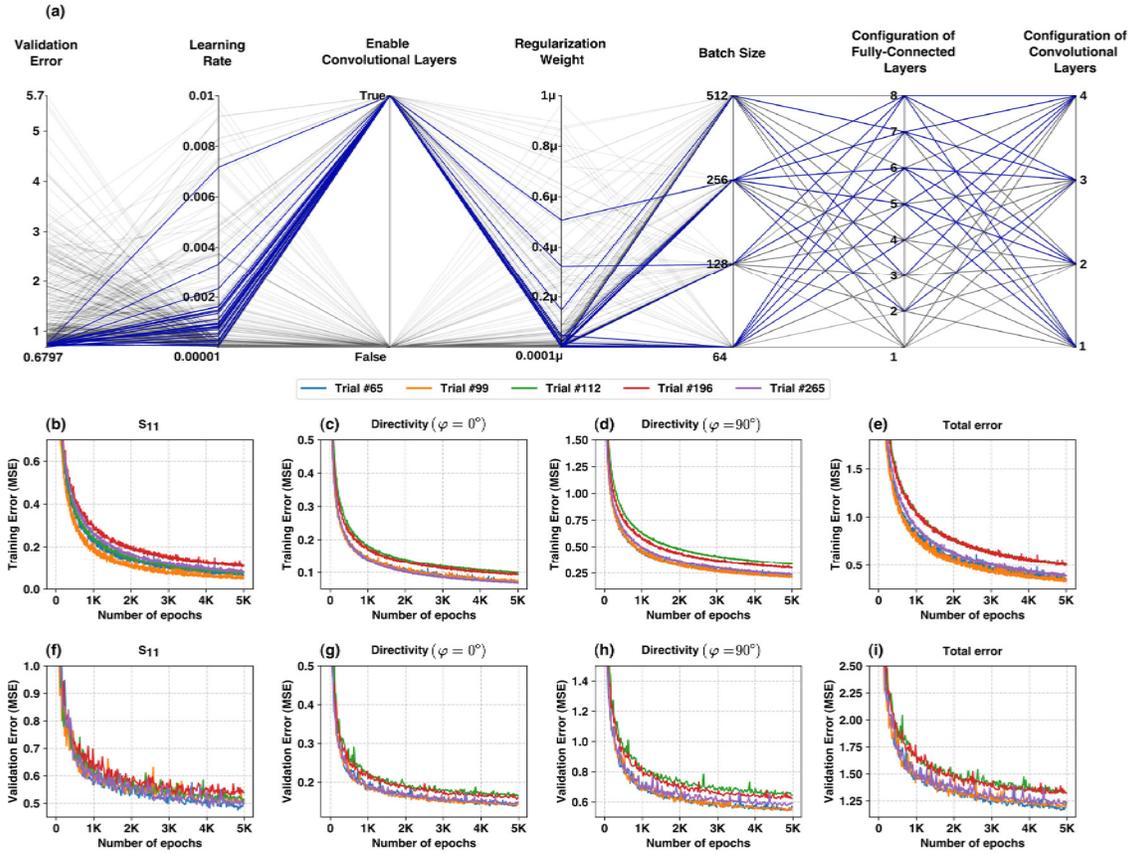

**Fig. 3.** (a) A parallel coordinate diagram shows the generated trials and their configuration during the Hyper-Parameter Optimization (HPO) process. The blue lines indicate the top 20 trials with the lowest validation error. (b-i) The training and validation learning curves of the top five trials generated by HPO. (b) training error of $S_{11}$, (c) training error of directivity ($\varphi = 0°$), (d) training error of directivity ($\varphi = 90°$), (e) total training error, (f) validation error of $S_{11}$, (g) validation error of directivity ($\varphi = 0°$), (h) validation error of directivity ($\varphi = 90°$), (i) total validation error.

To train the network during each trial, mean-squared error is used to measure the error between predicted and actual responses. To reduce the computational cost of this step, only the $S_{11}$ response is utilized during the hyperparameter optimization process, and the radiation pattern is employed only after the HPO trials. We use the Adam optimizer to learn the network weights, and the maximum number of epochs for the training of each trial is 1,000.

Figure 3(a) illustrates a parallel coordinate plot of the generated trials (300 trials have been generated in the HPO process), showing the sampled configurations and their corresponding validation errors. The diagram highlights the 20 trials with the lowest validation error ($S_{11}$), where all trials have a learning rate between 0.002 and $1e^{-5}$, use convolutional layers, and their regularization weight is less than $2e^{-7}$. We have selected the top five trials among the generated trials and trained them on the gathered dataset for longer epochs with the rest of responses. Figure 3(b)-(i) display the training and validation learning curves for this process. The curve indicates the successful convergence of models after 5,000 epochs without showing overfitting. We have selected the trial with the lowest validation error and evaluated the model on the test set to determine the model's overall accuracy and to ensure that the model does not overfit the validation set.

The selected architecture, depicted in Figure 1(d), comprises two main components: the backbone and the convolutional blocks. The backbone block takes the parameters of the device as the input and maps them into the latent space. This block comprises five fully connected layers, each with 512 neurons. The last fully connected layer is followed by three convolutional heads that map the features from the latent space to the response space. The responses predicted by each convolutional head are $S_{11}$ and radiation pattern in $\varphi = 0°$ and $\varphi = 90°$ planes, respectively. We have realized that a single head is enough to predict all the radiation patterns of different frequencies for each cut, due to the high linear correlation of the radiation patterns in the same cut. Since the radiation pattern changes gradually with frequency, a linear interpolation is utilized to approximate the radiation pattern at any frequency between 185 THz and 198.5 THz using the four approximated patterns from our surrogate solver. Given a device $d$, the trained surrogate solver is capable of estimating device responses $\hat{r} = \hat{f}(d)$, where $\hat{r}$ comprises both the estimated $S_{11}$ and radiation patterns in $\varphi = 0°$ and $\varphi = 90°$ planes, presented in concatenated vector form. Furthermore, the radiation patterns are interpolated at the specified frequency.

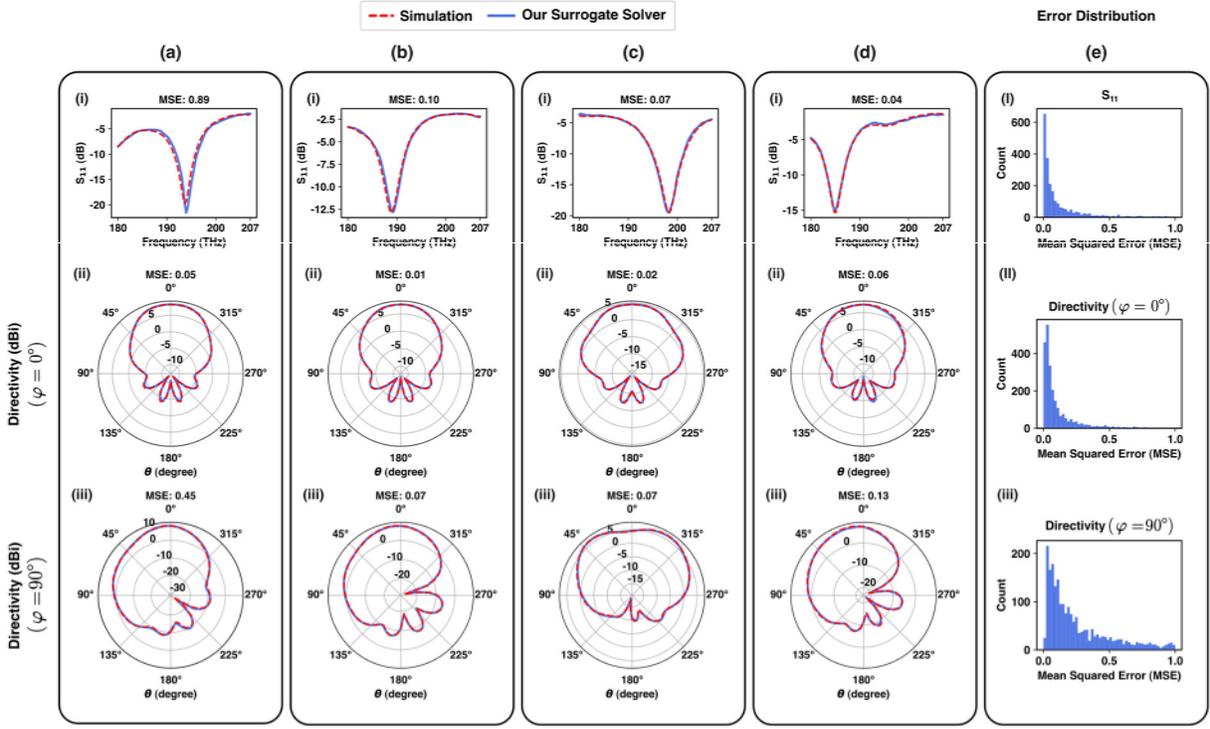

**Fig. 4.** The prediction accuracy and the error distribution of the proposed surrogate solver. (a-d) the simulated response and the predicted response of four devices, (e) the error distribution of each response type.

Due to the presence of spatial correlation in all responses, we designed the heads with convolutional layers instead of fully connected layers to effectively model spatial correlations and generate the final responses. Subsequently, during hyperparameter optimization, it was observed that convolutional layers consistently outperformed fully connected layers in capturing spatial correlations (see Supplementary section S4 for additional information regarding the configuration of the convolution blocks).

We have evaluated the overall accuracy of our network using the test set. The following are the mean squared errors for each response: the total error for $S_{11}$, the radiation pattern in $\varphi = 0°$, and $\varphi = 90°$ planes is, 0.53, 0.16, and 0.6, respectively. It is worth mentioning that the current antenna structures, have one plane of symmetry and the patterns are symmetric in $\varphi = 0°$ while non-symmetric in the $\varphi = 90°$ plane, leading to a higher error of the radiation pattern in $\varphi = 90°$ plane.

Figure 4(a)-(d) depict the qualitative prediction accuracy of the network. These figures show the simulated and predicted responses generated by the surrogate solver for four devices, demonstrating an almost perfect match between the two. To better illustrate the distribution of errors in each response, we have computed the error distribution plot as shown in Figure 4(e), where results indicate that the majority of the test samples exhibit errors < 1.0. More precisely, 90.84% of the samples have $S_{11}$ error < 1.0, 97.64% of the samples have directivity ($\varphi = 0°$) error < 1.0, and 85.68% of the samples have directivity ($\varphi = 90°$) error < 1.0. The numerical value of the MSE in each sample may not accurately reflect the quality of alignment between the predicted and target responses. This is due to the fact that, responses can have very large dips (for example, $S_{11}$ can have a dip value of -50 dB). A small difference between the predicted and target responses near the dip region may lead to a high squared error, despite the excellent visual alignment between the patterns, as shown in Figure 4. The MSE is primarily used for training purposes and to demonstrate the convergence of the process.

*2.4 Inverse Design Framework*

In our proposed framework, we have utilized the pseudo-inverse function to model the inverse problem. This approach offers several advantages, such as preserving one-to-many mappings, enabling the generation of multiple diverse designs (explained in 2.4.2), imposing fabrication constraints, and utilizing query-based objective functions (described in 2.4.3). The ultimate objective of our inverse design framework is to identify a set of candidate devices $D$ that satisfy the pseudo-inverse condition, given a desired response $r$:

$$D = \left\{d \in \mathbb{R}^k, \left\|\hat{f}(d) - r\right\|^2 < \varepsilon\right\}, \tag{3}$$

To achieve this, we have utilized the Particle Swarm Optimization (PSO) algorithm in combination with our neural network-based surrogate solver to efficiently explore the reduced device space ($\mathbb{R}^k$) and generate possible solutions for $D$. PSO is a population-based, meta-heuristic, evolutionary algorithm that is widely utilized in search and optimization problems and has proven as a highly effective approach for finding optimal solutions that minimize the objective function. The significant superiority of PSO over other alternatives such as genetic algorithm [69], apart from its straightforward implementation and accelerated convergence, lies in the memory retention of particles and the dynamic information exchange between them (information flow) [70], [71].

PSO starts by randomly creating particles to form a population in which each particle represents a unique configuration of a nanoantenna. In the next step, the population is evaluated using a predefined objective function (this evaluation is carried out simultaneously for all the particles using the surrogate solver). Consequently, particles are moved toward better solutions (with a lower objective function value) based on different factors, including the best local and global positions. The second and third steps are repeated iteratively until the particles are converged, and the results are used to determine a single optimum nanoantenna (Section 2.4.1) as well as multiple diverse nanoantennas (Section 2.4.2).

*2.4.1 Single Optimal Result*

To obtain a single optimal device that meets the desired response, the particle with the lowest objective function value is selected after PSO has converged. To evaluate the performance of our inverse method in generating single results, we have tasked our network with generating a single configuration of a nanoantenna for 2500 randomly sampled responses. The target $S_{11}$ is defined across all frequencies (180 THz - 198.5 THz), and the target radiation pattern is defined in $\varphi = 0°$ and $\varphi = 90°$ planes at four different frequencies. We have utilized the squared L2 distance as our objective function (discussed in 2.4.3), in which the entire shape of the generated and target responses must match. The target responses are extracted from the test set, consisting of randomly sampled devices with corresponding responses. In this experiment, we are confident that a device with the desired response exists within our design space. Our goal is to validate the capability of our inverse-design framework in finding these devices, also known as the physical targets [50], [51]. To quantitatively evaluate the performance of this experiment, we have calculated the mean squared error between the target and simulated responses of the generated devices. The total error for $S_{11}$, the radiation pattern in $\varphi = 0°$ and $\varphi = 90°$, is 0.46, 0.11, and 0.4, respectively. Figure 5 shows several instances of this evaluation where responses of the generated devices match very well with the target response (see Supplementary section S5, Figure S7 for the error distribution for this experiment).

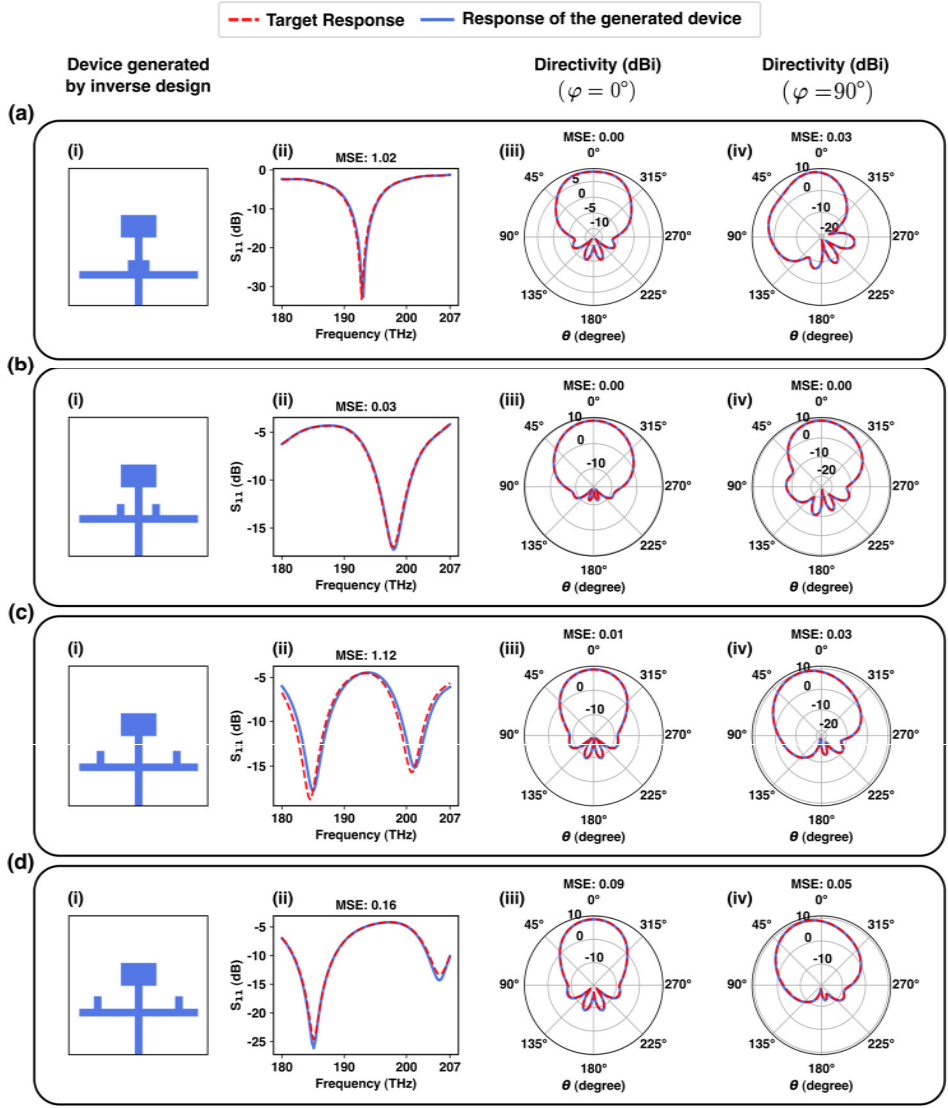

**Fig. 5.** Inverse design verification experiment with the goal of generating single optimal devices given the target responses. (a-d) the generated devices by the inverse design framework given the target responses.

*2.4.2 Multiple diverse results*

Due to the existence of one-to-many mappings in our dataset, a response may be realized with more than one device, creating several local minima in the optimization space. As PSO explores the design space, particles tend to get absorbed by regions where local minima are present. As a result, several clusters are formed after the convergence where the density of the particles is higher around local minima. In addition, using a query-based objective function brings flat regions into the optimization space, where several points meet the optimization criteria. To discover multiple diverse devices, the mean shift algorithm [72] is used to identify the clusters formed by PSO and locate the local minima. Mean shift is a non-parametric, density-based clustering algorithm used for segmentation and clustering, which can identify dense regions in the data space and finding local optima.

Figure 6(a)-(d) illustrate the procedure of discovering multiple diverse designs using the PSO particles and the mean shift algorithm. To obtain a set of diverse results $D$, after exploring the device space using a neural network-based surrogate solver, the resulting particles are filtered based on the value of the objective function, the clusters are identified using the mean-shift algorithm, and a set of candidate devices is determined by selecting the best particle in each cluster.

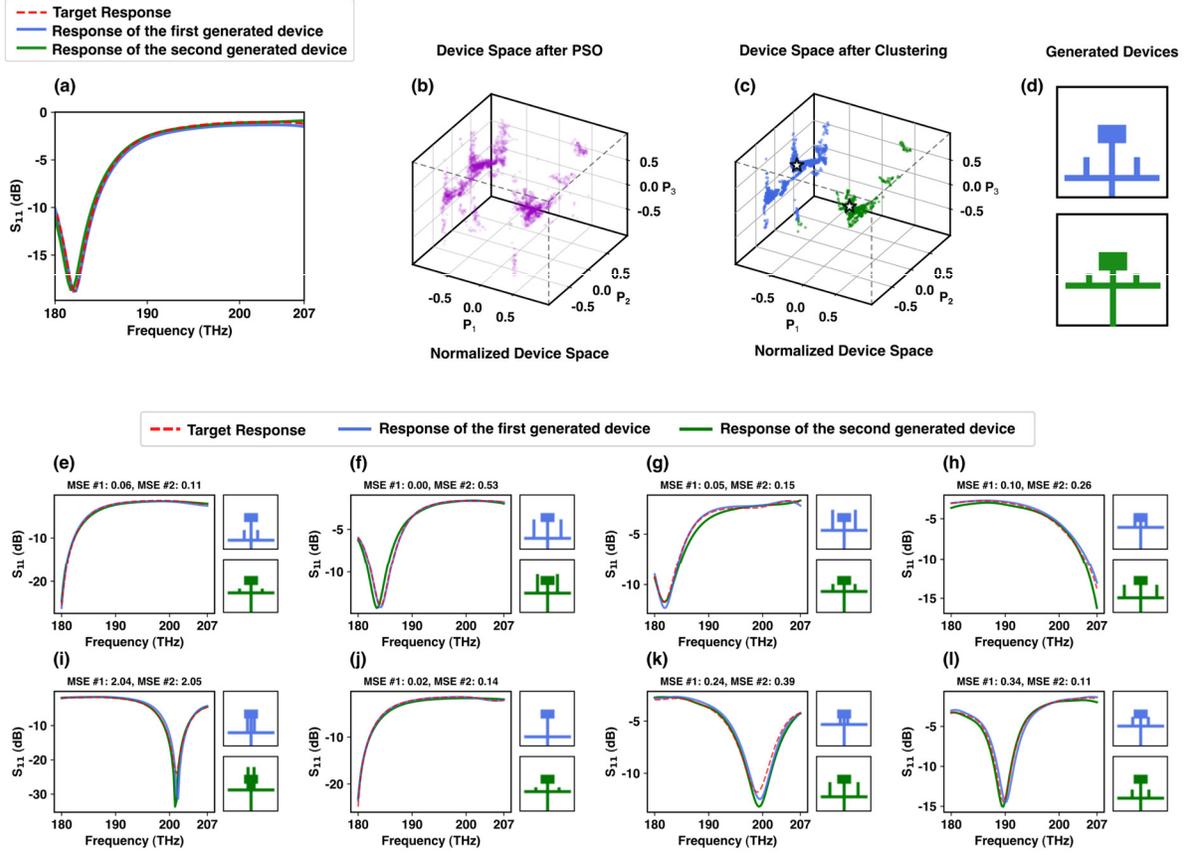

**Fig. 6.** (a-d) Examples of utilizing a clustering algorithm (mean shift) to generate multiple diverse devices given a target response. (a) The response of the generated devices and the target one, (b) device space after being explored by the PSO with the purple dots showing the particles, (c) particles are clustered into two groups using the mean shift algorithm, (d) generated devices by inverse design. (e-i) Qualitative examples of the experiment where the proposed inverse design framework was tasked to generate multiple diverse devices given the target response.

An experiment has been conducted to evaluate the accuracy of our inverse method in generating multiple diverse results with the same goal as in the previous section, where the squared L2 distance is used, and the generated response's shape must match the target response. However, in this evaluation, the inverse design framework is tasked with discovering more than one device for each target $S_{11}$. The qualitative results of this experiment are shown in Figures 6(e)-(i), indicating that the inverse design framework successfully discovered multiple configurations given the target responses. We have further improved the performance of PSO in exploring multiple local minima by prioritizing exploration over exploitation. This was achieved by increasing the number of particles and selecting the ring topology over fully connected. In the ring topology, particles can only communicate with their nearest neighbors, which limits their perception of the global minimum. This encourages distributed exploration in which the optimization space is partitioned into multiple regions, and particles operate exclusively within those specific areas.

*2.4.3 PSO's objective function*

Two different objective functions have been utilized throughout our framework for the PSO, the squared L2 distance and the query-based objective function. The squared L2 distance between the predicted and target responses is defined as follows:

$$l(d,r) = \left\|\hat{f}(d) - r\right\|^2, \tag{4}$$

where $r$ encapsulates the entire $S_{11}$ and the radiation pattern in the two radiation pattern cuts (at the specified frequency), all stacked together in a single vector. This type of objective function is useful when the generated response needs to precisely conform to the target response, i.e., to accurately match the target $S_{11}$ in the entire working frequency and the radiation pattern at every direction in the specified frequency. This objective function is used to evaluate the performance of the inverse design framework, quantitatively and qualitatively (Sections 2.5.1 and 2.5.2). It is also convenient to perform the inverse design task merely by defining a few conditions on the target response that the generated device must satisfy. This simplifies the inverse task for the designer, as one does not need to provide the full definition of the target response but only a few desired conditions. As a result, the framework can generate a wider range of candidate devices.

To achieve this, we have introduced and employed a query-based objective function. The term query refers to a request sent to the inverse-design framework to generate a nanoantenna. It contains a set of high-level conditions that the generated antenna should fulfill. For instance, to generate an efficient single band nanoantenna, the generated device must satisfy the following conditions: having an $S_{11}$ dip less than -10 dB at the antenna's working frequency and directivity more than 10 dBi in a specific direction. Each condition in the query is modeled as a cost function that takes the predicted response and returns an error scalar based on how well the requirement is met. The query-based objective function is then defined as a weighted sum of the cost functions, which is used as the objective function for the PSO:

$$l(d,r) = \sum_{c \in C} w_c c(\hat{f}(d), r), \qquad (5)$$

where $c$ is a cost function, $w_c$ is the corresponding weight, and $C$ is a set of cost functions of the specified query. We have used the query-based objective function to design a wide range of different nanoantennas which can be found in the results section.

In our framework, we can handle fabrication constraints in the following ways: Firstly, we can fix a set of device parameters in advance. For instance, we can task the framework to generate a device based on a desired response with a predefined stub location. Secondly, we can limit the range of motion of the parameters. For example, we can set a minimum and maximum value for the notch length. Finally, we can add an additional cost function to the PSO's objective function to account for any fabrication limitations. This will output a cost if the constraints are not met.

## 3  Results

In this section we will put our inverse design framework through a set of comprehensive tests to generate designs that not only satisfy the standard criteria required in real-world applications, but potentially outstripping them from performance point of view, while addressing complex design challenges. In this set of query-based experiments, we are unsure if the desired device exists in the design space, and whether it is physically possible to have a device with such responses (also known as non-physical targets [50], [51]). However, the inverse network generates the closest match that it can find in the design space.

*3.1 Single band nanoantennas with maximum directivity*

Typically, a single rectangular patch exhibits directivities in the range of 5-8 dBi; however, as shown in Figure 7(a), one can see that a directivity of up to 11.07 dBi is made possible for a single patch with our inverse design framework. In this section, the inverse design network has been tasked to design single band nanoantennas at frequencies $f$ = 193.5 THz with $S_{11}$ < -10 dB and the highest possible directivity in both $\varphi = 0°$ and $\varphi = 90°$ planes.

An interesting observation that can be made from the single band devices generated by our network (see Supplementary section S5 for more single band devices generated by our framework) is that lengths of the arms of the T-stubs are mostly short in length. This makes perfect sense from a physical point of view because if the length of the arm of the T-stub were longer, it would be either closer to the patch or pass through it. Either of the cases distorts the fringing field at the beginning side of the patch, acting as if there were slots in the patch, creating multiple resonances and making the patch act as multiple cavities (multi band operation). This is because a patch acts as a resonant cavity, and typically resonates when its physical dimensions correspond $\lambda/2$. Introducing slots into the patch modifies this resonant cavity by introducing discontinuities in the current distribution, changing the effective length of the antenna and distribution of the electric field across the patch, and the fringing fields, which is similar to creating multiple smaller resonators within the main patch (see Supplementary section S5 For another set of the single band antennas).

It should be noted that the radiation efficiency of all of the single band nanoantennas generated by our inverse-design framework is plotted in supplementary Figure S10(a), where radiation efficiencies of all of the nanoantennas lie approximately in the range of 70 – 75% at their central operational frequency, illustrating high radiation efficiencies, given the plasmonic nature of the structures. Additionally, polarization of the radiated waves, for all of the single band antennas, are illustrated in supplementary Figures S10(b-d) where the axial ratio which is the ratio of the major axis to the minor axis of the polarization ellipse is plotted against all $\varphi$ and $\theta$ angles in 2D equirectangular maps, showing linear polarization around the *z*-axis perpendicular to the antenna plane where the radiation is maximum.

In integrated circuits, planar antennas with half-space limited radiation patterns are of great interest, as they inherently prevent interference with the electronic and photonic components underneath them. Here, we will aim for the back lobe suppression, and the inverse design network has been tasked to design nanoantennas at frequencies $f$ = 185THz with the highest possible directivity at $\theta = 0°$, and minimum radiation at $\theta = 180°$, in both $\varphi = 0°$ and $\varphi = 90°$ planes. The generated device, and its responses are shown in Figure 7(b) (see Supplementary section S5 for more instances of back lobe-suppressed single band nanoantennas).

*3.2 Single band nanoantennas with constraints*

One of the major strengths of the proposed framework compared to its counterparts is that, apart from the fabrication/design constraints that were already applied in the dataset generation phase, further constraints can be applied to parameters after the training process. This is of great importance in cases where the training process is finished, however because of various design-specific, space-constraints, one wants to further limit the parameters. Typically, this process requires retraining the network again while considering these constraints, however in our framework, this can be done without retraining the network and simply by adding a set of constraints during the inverse process (explained in section 2.4.3).

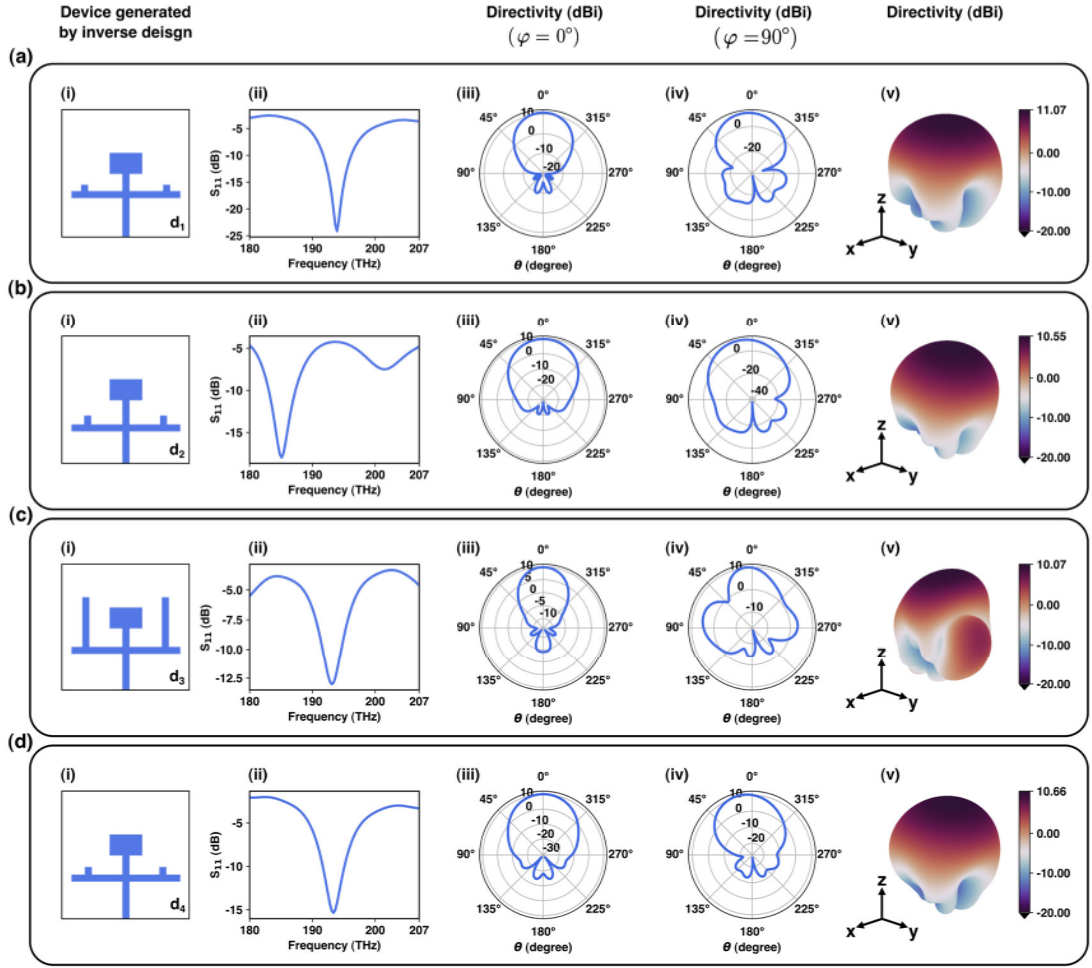

**Fig. 7.** Single band nanoantennas designed by the proposed inverse design framework: (a) a single band nanoantenna (device d1) designed for f = 193.5 THz with $S_{11}$ < -10 dB and the highest possible directivity in $\varphi = 0°$ and $\varphi = 90°$ planes. (b) a single band nanoantenna (device d2) designed for f = 185 THz with $S_{11}$ < -10 dB, the highest possible directivity in $\varphi = 0°$ and $\varphi = 90°$ planes, and a suppressed radiation in $\theta = 180°$. (c) and (d) show unconstrained and constrained single band nanoantennas (devices d3 and d4), respectively. Each subfigure (i-v) in panels (a-d) shows the schematic of the device, $S_{11}$, directivity in $\varphi = 0°$ plane, directivity in $\varphi = 90°$ plane, and the 3D radiation pattern for each of the devices, respectively.

As the first constraint, we will fix the length of the arm of the T-stub (as can be seen from a direct comparison between Figures 7(c) and 7(d)). This case is of particular importance in 2D arrays where the long length of the arm in the T-stub makes it difficult to have a dense array in the *y*-direction. To illustrate this point, we have first tasked the network to design an antenna with $S_{11}$

< -10 dB and directivity > 9 dBi at $f$ = 193.5 THz. Let us consider the cases where our network has generated devices with long T-stub arms as shown in Figure 7(c). As mentioned before, if there exist multiple devices that generate the same results (multiple clusters), our algorithm has the capability to choose either of the clusters (the network can be configured to either choose the best response, or any of the other clusters depending on the defined criteria). As a result, we will ask the network to only generate devices, satisfying the exact same queries at the same frequencies ($S_{11}$ < -10 dB and directivity > 9 dBi at $f$ = 193.5 THz), however this time, length of the arm of the T-stub is limited. The generated devices and their corresponding responses for the unconstrained and constrained cases are shown in Figures 7(c) and 7(d), respectively, perfectly illustrating the capabilities of our inverse network and the fact that, in order to impose constraints on the design, there is no need to retrain the network, thereby constraints can be applied even after the training process is finished (see Supplementary section S5 for the second case of imposing constraints where we will fix the location of the arm of the T-stub at 50 nm and the network is tasked to generate devices with $S_{11}$ < -10 dB and directivity > 8 dBi at $f$ = 186.5, 187.5, 189.5, 193.5, and 196.5 THz).

*3.3 Dual band and Broadband nanoantennas*

In this test, the network is tasked to generate dual band nanoantennas operating at two uncorrelated frequencies with $S_{11}$ < -10 dB and directivity > 8 dBi at both frequencies. We have tasked the network (as shown in Figures 8(a) and 8(b)) to specifically generate dual band antennas with two uncorrelated frequencies because uncorrelated frequencies do not share harmonics or other signal characteristics that can lead to cross-band interference, thereby they are less likely to interfere with each other. Moreover, since the two operational frequencies do not interfere with each other, they can be used simultaneously without degrading each other's performance, which leads to better spectrum efficiency. From the optical imaging point of view, dual band antennas can capture images at two different wavelengths simultaneously, providing richer information about the subject which is of huge interest in microscopy.

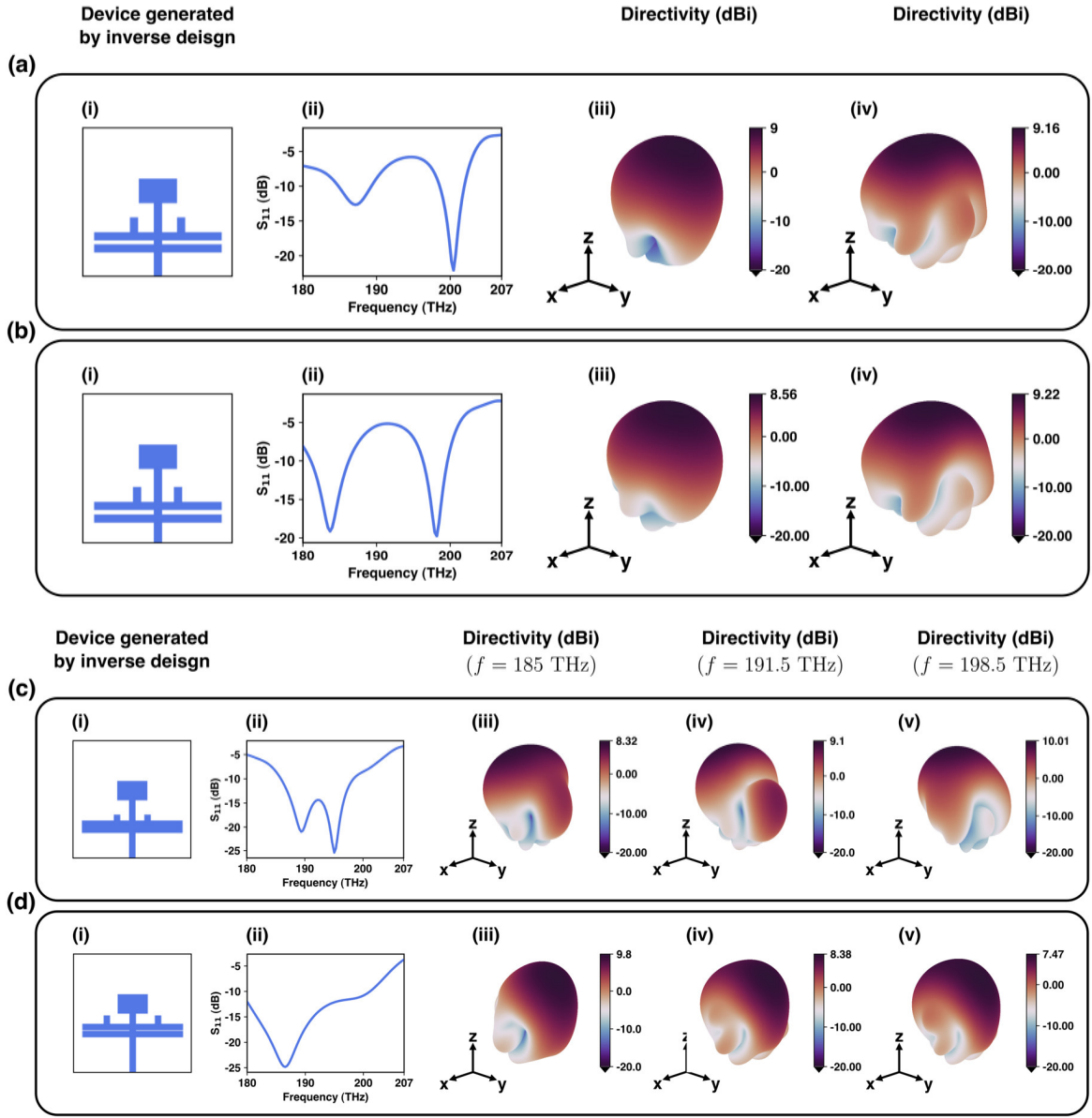

**Fig. 8.** Dual band and broadband nanoantennas designed by the proposed inverse design framework: (a) a dual band nanoantenna designed for f1 = 188 THz and f2 = 201 THz and (b) a dual band nanoantenna designed for f1 = 185 THz and f2 = 198.5 THz with $S_{11}$ < -10 dB and the highest possible directivity in $\varphi = 0°$ and $\varphi = 90°$ planes. (c) and (d) two broadband nanoantennas with bandwidths of 13.5 THz and more than 22 THz, respectively. Subfigures (i-iv) in panels (a) and (b) shows the schematic of the device, $S_{11}$, and directivities in the frequency dips of each device, respectively. Subfigures (i-v) in panels (c) and (d) shows the schematic of the device, $S_{11}$, and directivities in f = 185 THz, f = 191.5 THz, and f = 198.5 THz, respectively.

Additionally, since different wavelengths interact differently with various objects and materials, utilization of dual band antennas in LiDARs enables them to operate at two different wavelengths, leading to an improve in the resolution, accuracy, and better differentiation between different types of objects or materials.

Bandwidth plays a key role in the capacity of optical intra/inter-chip communication networks as broadband nanoantennas are capable of transmitting/receiving signals through multiple channels. As mentioned before, patch nanoantennas are inherently resonant structures and their impedance changes rapidly with frequency, leading to a large mismatch between the patch and the feed, resulting in their narrowband operation. As a result, having broadband plasmonic patch nanoantennas is of great importance in photonic integrated circuits due to their low-profile, planar nature. As the last test, we have tasked our network, as shown in Figure 8(c) and 8(d), to design broadband nanoantennas with $S_{11}$ < -10 dB and directivities > 8 dBi over the whole range. This superior broadband feature while maintaining relatively high directivities over the whole range, enables the plasmonic patch nanoantennas to play a pivotal role in photonic integrated circuits.

*3.4 Multiple-diverse results*

In this section, our inverse design framework is tasked to generate three different devices for each set of defined criteria, shown in Figures 10(a)-(c), respectively. For instance, Figure 9(a) illustrates three different devices that have been generated by our inverse-design framework with $S_{11}$ < -10 dB, and directivity > 10 dBi in both $\varphi = 0°$ and $\varphi = 90°$ planes, at $f$ = 187.5 THz. Figures 10(b) and 10(c) also show three different devices generated by our inverse network, with the same criteria, but this time at $f$ = 193.5 THz and $f$ = 198.5 THz. As it is obvious in Figure 9, the proposed framework is capable of successfully generating multiple devices for a set of criteria, each of which can be used for different applications and according to different design limitations.

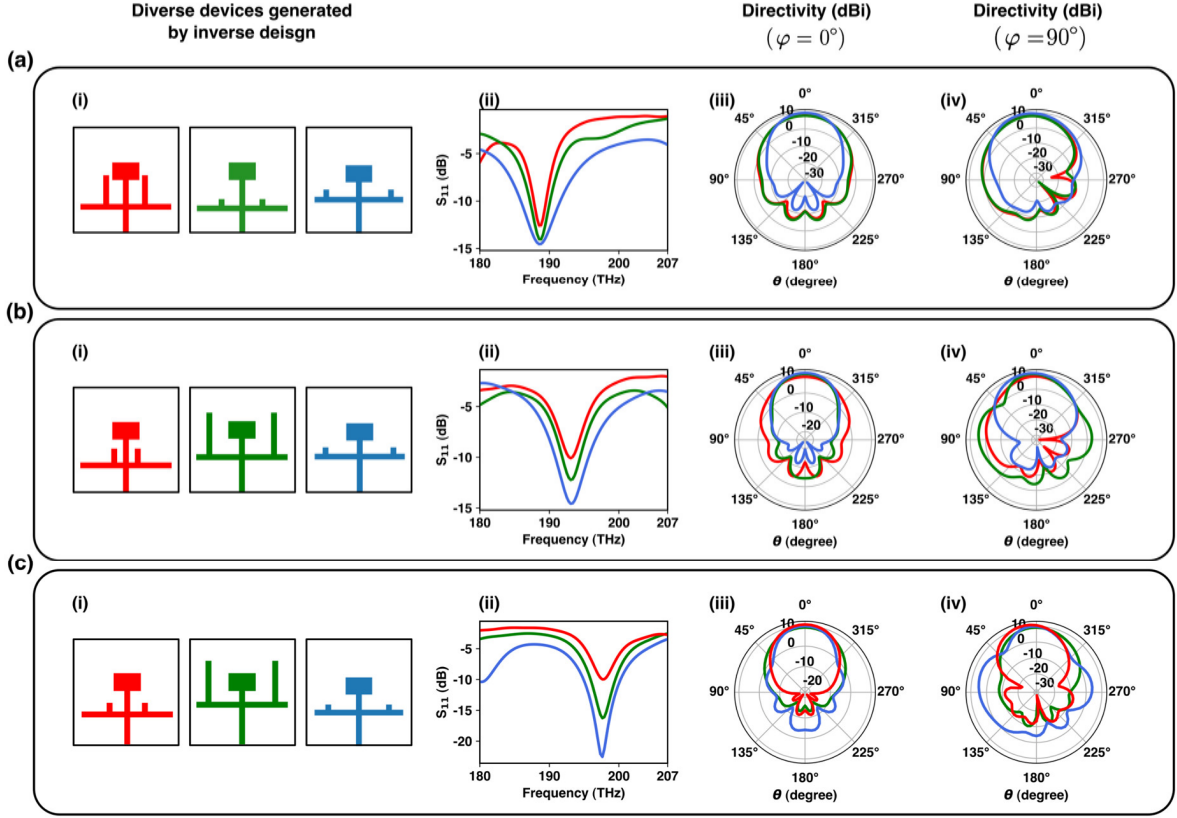

**Fig. 9.** (a)-(c) multiple diverse devices generated for the same query. Sub-figures (i-iv) in (a)-(d) represent (i) three different nanoantennas generated for the same query, (ii) their corresponding $S_{11}$, (iii) directivity in phi and (iv) directivity in phi = 90 planes, respectively. Red, Green, and Blue, plots in each subpanel (ii-iv) correspond to the nanoantenna of the same colour.

## 4    Discussion

In this section we will discuss the nonlinearity of the current problem, and how the proposed method can be extended for use in inverse design of random media where extremely large datasets are required.

*4.1  Discussion on nonlinearity of the problem*

Although the parameter space (3 and 4 parameters) may seem small at the first sight, the amount of nonlinearity they introduce to the result space is significant. In our work, the combination of stubs and their arms serve as the impedance matching network, either for single-band, dual-band, or broadband cases. The nonlinearity of the problem is partially embedded in the context of impedance matching and the fact that the input impedance of each section with a length $l$ shows a $\tan(\beta l)$ or $\cot(\beta l)$ dependency, which itself is the source of nonlinearity of the problem because of the nonlinear behavior of the tangent function near its poles (in various certain regions a slight change in a parameter leads to a huge change in the response). On the other hand, addition of the arm on a stub introduces another level of frequency sensitivity: at the junction of the arm and the stub the input impedance depends on both $\tan(\beta l_1)$ and $\tan(\beta l_2)$ and as a result, location of the arm and its length introduces a relatively significant amount of nonlinearity in certain regions of the input impedance function at the junction of the stub and its arm (see supplementary Figures S13), showing two examples of this nonlinearity, where a rather slight change in the location of the arm of the T-stub ($D_a$) leads to emergence of nonlinear changes in the $S_{11}$). Furthermore, adding the second stub (as in the 4-parameter case) results in a new impedance transformation path on the Smith chart which increases the nonlinearity (since with combination of two stubs, the impedance point can traverse a potentially looping or crisscrossing through different reactance and resistance levels paths). All of the mentioned nonlinearities, combine with another significant source of nonlinearity of the problem which is the coupling between different sections of the structure such as the coupling between stubs, T-stub's arm, and the patch.

*4.2 Novelty of the current framework and its potential for extension to problem consisting random media*

It is important to briefly mention the novelties of the framework and why the proposed approach is a good candidate for use in random media, such as deep tissue imaging [73], random lasers [74], study of coherent backscattering [75], [76], quantum information processing [77], and random metasurfaces [78].

Non-uniqueness nature of the inverse problem where a response can be realized by multiple devices, makes it challenging to directly learn the inverse mapping using discriminative neural networks. Approaches, such as tandem networks [58], aim to reduce the one-to-many mappings to a one-to-one mapping in order to learn the inverse mapping directly. However, this may result in eliminating devices from the device space that could still be useful. It is important to generate multiple devices because it allows the designer to choose a device that exhibits less sensitivity to fabrication imperfections. When two devices have the same desired response, the one that is more stable and less sensitive to fabrication is preferred. Generative approaches, including those based on variational autoencoders [60], [61] and generative adversarial networks [43], [62], [63], [64], have the ability to generate multiple devices. However, due to the complexity of training, these approaches may still suffer from mode collapses and fail to adequately capture the diversity of the device space.

An important capability of our network that renders very useful for inverse design of random media is preserving the one-to-many mappings, which is crucial for capturing the inherent complexity of the problem as it enriches the dataset for training of the network, leads to better generalization and prediction capabilities, and facilitates an effective and comprehensive exploration of the design space, leading to discovery of optimal solutions that might otherwise be overlooked. Additionally, many of the output devices may not be exactly realizable due to various factors (such as environmental variations in real-life scenarios, special arrangements of scatterers that might be hard to fabricate, etc.), as a result preserving the one-to-many mappings will handle this issue effectively by offering multiple alternative solutions. Moreover, in real-life random media scenarios, measurement noise and statistical fluctuations in the arrangement of scatterers, can lead to discrepancies in the design process and performance. Preserving the one-to-many mappings mitigates the impact of such noises/fluctuations by offering alternative solutions.

Additionally, the proposed approach allows for adjustments and optimizations by applying post-training constraints, which is of great importance, especially in random media where many of the output devices may not be fabricable and the applications of further constraints are mandatory (which can be done without any retraining procedure using our network). This means that if a device with a desired response exhibits different, unwanted behavior after fabrication due to errors in the fabrication process (for example, the feed and arm of the T-stub being merged due to fabrication inaccuracy), we can ask the framework to find another device with the same response but less sensitive to the fabrication imperfections. For instance, this can be achieved by fixing a parameter in the PSO search space, limiting the range of motion of the parameters, or by defining an additional cost function in the PSO objective function. In tandem networks [58], one-to-many mappings are eliminated, meaning that for each response, only one device can fulfill it, making it impossible to have other devices that meet the fabrication constraints. In generative networks also, this aspect has not been explored, especially in approaches with larger degrees of freedom that generate freeform objects, leading to devices that are not fabricable.

The proposed architecture has been developed to learn from smaller dataset sizes and presents a relatively good level of generalization (considering the nonlinearity of the problem mentioned earlier in section 4.1), is a good candidate for utilization in inverse design problems concerning random media with a large number of parameters, where extremely large datasets are required for the network to learn the relationships in those highly nonlinear spaces.

It is important to note that in generative approaches and tandem networks, the inverse function $f^{-1}: r \to d$ is directly modeled, where the entire response $r \in \mathbb{R}^{672}$ is required to be provided to generate a device. However, this may not be favorable as only the response in a specific region $r \in \mathbb{R}^q$ might be of interest ($q \ll 672$), and providing arbitrary responses in the rest of the regions may limit the diversity of the generated devices to those that exactly exhibit $r \in \mathbb{R}^{672}$. In our framework, we can use a query-based approach to search for the desired device simply by defining a few conditions instead of providing the entire response. This results in finding additional devices that exhibit the desired behavior.

### 4.3 Importance of using PSO as the search algorithm

PSO and GA [69] are the most prominent optimization algorithms used extensively in numerous applications due to their versatility, robustness, and their ability to navigate through complex, high dimensional problems. Apart from PSO's simplicity, fewer tuning parameters, and faster convergence speed in rugged and complex spaces, in the following, we will discuss multiple reasonings behind the choice of PSO over GA for our network.

The most significant advantage of PSO over other evolutionary algorithms, such as GA, lies in the inherent memory of the particles and the information flow between them. Each particle in PSO has a memory and shares its own experience with all other particles while obeying universal rules. Consequently, each particle benefits from other particles' knowledge, enabling the swarm to efficiently navigate through complex and rugged spaces collectively. Moreover, PSO does not rely on gradient information, making it a good candidate for problems with highly nonlinear, non-differentiable, or noisy objective functions.

In PSO, each particle's position is updated based on its personal experience and its knowledge of the global experience shared by other particles, which will result in a balance between exploration and exploitation by preventing premature convergence in highly nonlinear spaces. This is in a strong contrast with GA, which relies on random discrete crossover and mutation operators and struggles to maintain this balance, leading to over-exploitation or insufficient exploration of the search space. This is due to the fact that crossover operators may produce offsprings without inheriting the useful features of parents, leading to poor exploration whereas mutation often comes at the cost of disrupting the existing good solutions, making it very challenging to find the optimum(s) in problems with rugged spaces. The discreteness of the crossover and mutation operators in GA also introduces abrupt changes in the search space and may lead to fluctuations in the errors, contrary to PSO, which benefits from a smoother search trajectory due to the continuous nature of adjustment of particle velocities.

Furthermore, over successive generations, GAs tend to lose population diversity, especially in cases where selection pressure is high. This loss of diversity mitigates the algorithm's ability to escape local minima, which is more pronounced in highly nonlinear problems.

**Conclusion**

In this study, we have developed an efficient framework for the inverse design of plasmonic patch nanoantennas in the NIR regime. Our framework can design a wide range of devices including single band, dual band, and broadband antennas, with directivities of up to 11.07 dBi and radiation efficiencies reaching 75% for a single patch. Moreover, our approach demonstrates a remarkable versatility in terms of applying various post-training design and application-specific constraints where, in addition to the primary fabrication constrains that have been considered while generating the dataset, further constraints can also be applied after the training process. This is crucial in addressing the ever-expanding needs of modern optical phased arrays, where designers are dealing with increasingly strictly stringent integration requirements. The proposed approach preserves the one-to-many mappings and provides the designer with the ability to choose from multiple diverse designs, given specific geometry and constraints. Our approach takes a significant departure from traditional NN-based inverse-design methods and sets a precedent for future research in the field leveraging the robust predictive and generative capabilities of deep neural networks in optical designs. This paradigm shift towards an inverse design approach fosters a more efficient and creative design process, enabling the exploration of innovative optical designs that might be overlooked or infeasible in conventional forward and inverse-design methods.

**References**

[1] E. Yablonovitch, "Inhibited Spontaneous Emission in Solid-State Physics and Electronics," *Phys. Rev. Lett.*, vol. 58, no. 20, pp. 2059–2062, May 1987, doi: 10.1103/PhysRevLett.58.2059.

[2] S. Y. Lin *et al.*, "A three-dimensional photonic crystal operating at infrared wavelengths," *Nature*, vol. 394, no. 6690, pp. 251–253, Jul. 1998, doi: 10.1038/28343.

[3] J. B. Pendry, D. Schurig, and D. R. Smith, "Controlling Electromagnetic Fields," *Science*, vol. 312, no. 5781, pp. 1780–1782, Jun. 2006, doi: 10.1126/science.1125907.

[4] D. R. Smith, J. B. Pendry, and M. C. K. Wiltshire, "Metamaterials and Negative Refractive Index," *Science*, vol. 305, no. 5685, pp. 788–792, Aug. 2004, doi: 10.1126/science.1096796.

[5] N. Yu *et al.*, "Light Propagation with Phase Discontinuities: Generalized Laws of Reflection and Refraction," *Science*, vol. 334, no. 6054, pp. 333–337, Oct. 2011, doi: 10.1126/science.1210713.

[6] M. Khorasaninejad, W. T. Chen, R. C. Devlin, J. Oh, A. Y. Zhu, and F. Capasso, "Metalenses at visible wavelengths: Diffraction-limited focusing and subwavelength resolution imaging," *Science*, vol. 352, no. 6290, pp. 1190–1194, Jun. 2016, doi: 10.1126/science.aaf6644.

[7] L. Hsu, M. Dupré, A. Ndao, and B. Kanté, "From parabolic-trough to metasurface-concentrator: assessing focusing in the wave-optics limit," *Opt. Lett.*, vol. 42, no. 8, p. 1520, Apr. 2017, doi: 10.1364/OL.42.001520.

[8] J. Ha, A. Ndao, L. Hsu, J.-H. Park, and B. Kante, "Planar dielectric cylindrical lens at 800 nm and the role of fabrication imperfections," *Opt. Express*, vol. 26, no. 18, p. 23178, Sep. 2018, doi: 10.1364/OE.26.023178.

[9]  A. Ndao, L. Hsu, J. Ha, J.-H. Park, C. Chang-Hasnain, and B. Kanté, "Octave bandwidth photonic fishnet-achromatic-metalens," *Nat Commun*, vol. 11, no. 1, p. 3205, Dec. 2020, doi: 10.1038/s41467-020-17015-9.

[10] L. Hsu and A. Ndao, "Diffraction-limited broadband optical meta-power-limiter," *Opt. Lett.*, vol. 46, no. 6, p. 1293, Mar. 2021, doi: 10.1364/OL.418745.

[11] W. T. Chen *et al.*, "A broadband achromatic metalens for focusing and imaging in the visible," *Nature Nanotech*, vol. 13, no. 3, pp. 220–226, Mar. 2018, doi: 10.1038/s41565-017-0034-6.

[12] S. Moayed Baharlou, S. Hemayat, K. C. Toussaint, and A. Ndao, "GPU-Accelerated and Memory-Independent Layout Generation for Arbitrarily Large-Scale Metadevices," *Advcd Theory and Sims*, vol. 7, no. 1, p. 2300378, Jan. 2024, doi: 10.1002/adts.202300378.

[13] M. W. Khalid *et al.*, "Meta-Magnetic All-Optical Helicity Dependent Switching of Ferromagnetic Thin Films," *Advanced Optical Materials*, p. 2301599, Oct. 2023, doi: 10.1002/adom.202301599.

[14] W. T. Chen *et al.*, "Generation of wavelength-independent subwavelength Bessel beams using metasurfaces," *Light Sci Appl*, vol. 6, no. 5, p. e16259, May 2017, doi: 10.1038/lsa.2016.259.

[15] S. Hemayat, L. Hsu, J. Ha, and A. Ndao, "Near-unity uniformity and efficiency broadband meta-beam-splitter/combiner," *Opt. Express*, vol. 31, no. 3, p. 3984, Jan. 2023, doi: 10.1364/OE.480233.

[16] J. A. Schuller, E. S. Barnard, W. Cai, Y. C. Jun, J. S. White, and M. L. Brongersma, "Plasmonics for extreme light concentration and manipulation," *Nature Mater*, vol. 9, no. 3, pp. 193–204, Mar. 2010, doi: 10.1038/nmat2630.

[17] S. A. Maier, *Plasmonics: Fundamentals and Applications*. New York, NY: Springer US, 2007. doi: 10.1007/0-387-37825-1.

[18] J.-H. Park *et al.*, "Symmetry-breaking-induced plasmonic exceptional points and nanoscale sensing," *Nat. Phys.*, vol. 16, no. 4, pp. 462–468, Apr. 2020, doi: 10.1038/s41567-020-0796-x.

[19] J.-H. Park, A. Kodigala, A. Ndao, and B. Kanté, "Hybridized metamaterial platform for nano-scale sensing," *Opt. Express*, vol. 25, no. 13, p. 15590, Jun. 2017, doi: 10.1364/OE.25.015590.

[20] L. Hsu, F. I. Baida, and A. Ndao, "Local field enhancement using a photonic-plasmonic nanostructure," *Opt. Express*, vol. 29, no. 2, p. 1102, Jan. 2021, doi: 10.1364/OE.415956.

[21] W. Srituravanich, L. Pan, Y. Wang, C. Sun, D. B. Bogy, and X. Zhang, "Flying plasmonic lens in the near field for high-speed nanolithography," *Nature Nanotech*, vol. 3, no. 12, pp. 733–737, Dec. 2008, doi: 10.1038/nnano.2008.303.

[22] Z. Liu, J. M. Steele, W. Srituravanich, Y. Pikus, C. Sun, and X. Zhang, "Focusing Surface Plasmons with a Plasmonic Lens," *Nano Lett.*, vol. 5, no. 9, pp. 1726–1729, Sep. 2005, doi: 10.1021/nl051013j.

[23] K. Wang, E. Schonbrun, P. Steinvurzel, and K. B. Crozier, "Trapping and rotating nanoparticles using a plasmonic nano-tweezer with an integrated heat sink," *Nat Commun*, vol. 2, no. 1, p. 469, Sep. 2011, doi: 10.1038/ncomms1480.

[24] K. B. Crozier, "Quo vadis, plasmonic optical tweezers?," *Light Sci Appl*, vol. 8, no. 1, p. 35, Dec. 2019, doi: 10.1038/s41377-019-0146-x.

[25] S. Hemayat and S. Darbari, "Far-field position-tunable trapping of dielectric particles using a graphene-based plasmonic lens," *Opt. Express*, vol. 30, no. 4, p. 5512, Feb. 2022, doi: 10.1364/OE.451740.

[26] L. Yousefi and A. C. Foster, "Waveguide-fed optical hybrid plasmonic patch nano-antenna," *Opt. Express*, vol. 20, no. 16, p. 18326, Jul. 2012, doi: 10.1364/OE.20.018326.

[27] B. A. Nia, L. Yousefi, and M. Shahabadi, "Integrated Optical-Phased Array Nanoantenna System Using a Plasmonic Rotman Lens," *J. Lightwave Technol.*, vol. 34, no. 9, pp. 2118–2126, May 2016, doi: 10.1109/JLT.2016.2520881.

[28] G. Kaplan, K. Aydin, and J. Scheuer, "Dynamically controlled plasmonic nano-antenna phased array utilizing vanadium dioxide [Invited]," *Opt. Mater. Express*, vol. 5, no. 11, p. 2513, Nov. 2015, doi: 10.1364/OME.5.002513.

[29] X. Ni, N. K. Emani, A. V. Kildishev, A. Boltasseva, and V. M. Shalaev, "Broadband Light Bending with Plasmonic Nanoantennas," *Science*, vol. 335, no. 6067, pp. 427–427, Jan. 2012, doi: 10.1126/science.1214686.

[30] L. Huang *et al.*, "Three-dimensional optical holography using a plasmonic metasurface," *Nat Commun*, vol. 4, no. 1, p. 2808, Nov. 2013, doi: 10.1038/ncomms3808.

[31] N. Palombo Blascetta *et al.*, "Nanoscale Imaging and Control of Hexagonal Boron Nitride Single Photon Emitters by a Resonant Nanoantenna," *Nano Lett.*, vol. 20, no. 3, pp. 1992–1999, Mar. 2020, doi: 10.1021/acs.nanolett.9b05268.

[32] Z. Zhu, B. Bai, O. You, Q. Li, and S. Fan, "Fano resonance boosted cascaded optical field enhancement in a plasmonic nanoparticle-in-cavity nanoantenna array and its SERS application," *Light Sci Appl*, vol. 4, no. 6, pp. e296–e296, Jun. 2015, doi: 10.1038/lsa.2015.69.

[33] G. S. Unal and M. I. Aksun, "Bridging the Gap between RF and Optical Patch Antenna Analysis via the Cavity Model," *Sci Rep*, vol. 5, no. 1, p. 15941, Nov. 2015, doi: 10.1038/srep15941.

[34] A. Krizhevsky, I. Sutskever, and G. E. Hinton, "ImageNet Classification with Deep Convolutional Neural Networks," in *Advances in Neural Information Processing Systems*,

F. Pereira, C. J. Burges, L. Bottou, and K. Q. Weinberger, Eds., Curran Associates, Inc., 2012. [Online]. Available: https://proceedings.neurips.cc/paper_files/paper/2012/file/c399862d3b9d6b76c8436e924a68c45b-Paper.pdf

[35] K. He, X. Zhang, S. Ren, and J. Sun, "Deep Residual Learning for Image Recognition," in *2016 IEEE Conference on Computer Vision and Pattern Recognition (CVPR)*, Las Vegas, NV, USA: IEEE, Jun. 2016, pp. 770–778. doi: 10.1109/CVPR.2016.90.

[36] I. Sutskever, O. Vinyals, and Q. V. Le, "Sequence to Sequence Learning with Neural Networks," in *Proceedings of the 27th International Conference on Neural Information Processing Systems - Volume 2*, in NIPS'14. Cambridge, MA, USA: MIT Press, 2014, pp. 3104–3112.

[37] A. Vaswani *et al.*, "Attention is All You Need," in *Proceedings of the 31st International Conference on Neural Information Processing Systems*, in NIPS'17. Red Hook, NY, USA: Curran Associates Inc., 2017, pp. 6000–6010.

[38] G. Hinton *et al.*, "Deep Neural Networks for Acoustic Modeling in Speech Recognition: The Shared Views of Four Research Groups," *IEEE Signal Process. Mag.*, vol. 29, no. 6, pp. 82–97, Nov. 2012, doi: 10.1109/MSP.2012.2205597.

[39] I. Malkiel, M. Mrejen, A. Nagler, U. Arieli, L. Wolf, and H. Suchowski, "Plasmonic nanostructure design and characterization via Deep Learning," *Light Sci Appl*, vol. 7, no. 1, p. 60, Sep. 2018, doi: 10.1038/s41377-018-0060-7.

[40] C. C. Nadell, B. Huang, J. M. Malof, and W. J. Padilla, "Deep learning for accelerated all-dielectric metasurface design," *Opt. Express*, vol. 27, no. 20, p. 27523, Sep. 2019, doi: 10.1364/OE.27.027523.

[41] J. Peurifoy *et al.*, "Nanophotonic particle simulation and inverse design using artificial neural networks," *Sci. Adv.*, vol. 4, no. 6, p. eaar4206, Jun. 2018, doi: 10.1126/sciadv.aar4206.

[42] P. R. Wiecha and O. L. Muskens, "Deep Learning Meets Nanophotonics: A Generalized Accurate Predictor for Near Fields and Far Fields of Arbitrary 3D Nanostructures," *Nano Lett.*, vol. 20, no. 1, pp. 329–338, Jan. 2020, doi: 10.1021/acs.nanolett.9b03971.

[43] S. So and J. Rho, "Designing nanophotonic structures using conditional deep convolutional generative adversarial networks," *Nanophotonics*, vol. 8, no. 7, pp. 1255–1261, Jul. 2019, doi: 10.1515/nanoph-2019-0117.

[44] J. Noh *et al.*, "Design of a transmissive metasurface antenna using deep neural networks," *Opt. Mater. Express*, vol. 11, no. 7, pp. 2310–2317, Jul. 2021, doi: 10.1364/OME.421990.

[45] S. So, D. Lee, T. Badloe, and J. Rho, "Inverse design of ultra-narrowband selective thermal emitters designed by artificial neural networks," *Opt. Mater. Express*, vol. 11, no. 7, pp. 1863–1873, Jul. 2021, doi: 10.1364/OME.430306.

[46] J. Noh *et al.*, "Reconfigurable reflective metasurface reinforced by optimizing mutual coupling based on a deep neural network," *Photonics and Nanostructures - Fundamentals and Applications*, vol. 52, p. 101071, 2022, doi: https://doi.org/10.1016/j.photonics.2022.101071.

[47] S. So, J. Mun, and J. Rho, "Simultaneous Inverse Design of Materials and Structures via Deep Learning: Demonstration of Dipole Resonance Engineering Using Core–Shell Nanoparticles," *ACS Applied Materials & Interfaces*, vol. 11, no. 27, pp. 24264–24268, 2019, doi: 10.1021/acsami.9b05857.

[48] S. So, T. Badloe, J. Noh, J. Bravo-Abad, and J. Rho, "Deep learning enabled inverse design in nanophotonics," *Nanophotonics*, vol. 9, no. 5, pp. 1041–1057, 2020, doi: doi:10.1515/nanoph-2019-0474.

[49] W. Li *et al.*, "Machine Learning for Engineering Meta-Atoms with Tailored Multipolar Resonances," *Laser & Photonics Reviews*, vol. n/a, no. n/a, p. 2300855, doi: https://doi.org/10.1002/lpor.202300855.

[50] A. Estrada-Real, A. Khaireh-Walieh, B. Urbaszek, and P. R. Wiecha, "Inverse design with flexible design targets via deep learning: Tailoring of electric and magnetic multipole scattering from nano-spheres," *Photonics and Nanostructures - Fundamentals and Applications*, vol. 52, p. 101066, 2022, doi: https://doi.org/10.1016/j.photonics.2022.101066.

[51] A. Vallone, N. M. Estakhri, and N. M. Estakhri, "Region-specified inverse design of absorption and scattering in nanoparticles by using machine learning," *Journal of Physics: Photonics*, vol. 5, no. 2, p. 024002, Apr. 2023, doi: 10.1088/2515-7647/acc7e5.

[52] D. Gostimirovic, Y. Grinberg, D.-X. Xu, and O. Liboiron-Ladouceur, "Improving Fabrication Fidelity of Integrated Nanophotonic Devices Using Deep Learning," *ACS Photonics*, vol. 10, no. 6, pp. 1953–1961, Jun. 2023, doi: 10.1021/acsphotonics.3c00389.

[53] O. Buchnev, J. A. Grant-Jacob, R. W. Eason, N. I. Zheludev, B. Mills, and K. F. MacDonald, "Deep-Learning-Assisted Focused Ion Beam Nanofabrication," *Nano Lett.*, vol. 22, no. 7, pp. 2734–2739, Apr. 2022, doi: 10.1021/acs.nanolett.1c04604.

[54] D. Melati *et al.*, "Mapping the global design space of nanophotonic components using machine learning pattern recognition," *Nat Commun*, vol. 10, no. 1, p. 4775, Oct. 2019, doi: 10.1038/s41467-019-12698-1.

[55] Y. Liu, T. Lu, K. Wu, and J.-M. Jin, "A Hybrid Algorithm for Electromagnetic Optimization Utilizing Neural Networks," in *2018 IEEE 27th Conference on Electrical Performance of Electronic Packaging and Systems (EPEPS)*, San Jose, CA: IEEE, Oct. 2018, pp. 261–263. doi: 10.1109/EPEPS.2018.8534264.

[56] Z. Ma and Y. Li, "Parameter extraction and inverse design of semiconductor lasers based on the deep learning and particle swarm optimization method," *Opt. Express*, vol. 28, no. 15, pp. 21971–21981, Jul. 2020, doi: 10.1364/OE.389474.

[57] C. Zhang, G. Kang, J. Wang, Y. Pan, and J. Qu, "Inverse design of soliton microcomb based on genetic algorithm and deep learning," *Opt. Express*, vol. 30, no. 25, pp. 44395–44407, Dec. 2022, doi: 10.1364/OE.471706.

[58] D. Liu, Y. Tan, E. Khoram, and Z. Yu, "Training Deep Neural Networks for the Inverse Design of Nanophotonic Structures," *ACS Photonics*, vol. 5, no. 4, pp. 1365–1369, Apr. 2018, doi: 10.1021/acsphotonics.7b01377.

[59] J. H. Han *et al.*, "Neural-Network-Enabled Design of a Chiral Plasmonic Nanodimer for Target-Specific Chirality Sensing," *ACS Nano*, vol. 17, no. 3, pp. 2306–2317, Feb. 2023, doi: 10.1021/acsnano.2c08867.

[60] D. P. Kingma and M. Welling, "Auto-Encoding Variational Bayes," 2013, doi: 10.48550/ARXIV.1312.6114.

[61] W. Ma, F. Cheng, Y. Xu, Q. Wen, and Y. Liu, "Probabilistic Representation and Inverse Design of Metamaterials Based on a Deep Generative Model with Semi-Supervised Learning Strategy," *Advanced Materials*, vol. 31, no. 35, p. 1901111, Aug. 2019, doi: 10.1002/adma.201901111.

[62] I. J. Goodfellow *et al.*, "Generative Adversarial Networks," 2014, doi: 10.48550/ARXIV.1406.2661.

[63] J. Jiang, D. Sell, S. Hoyer, J. Hickey, J. Yang, and J. A. Fan, "Free-Form Diffractive Metagrating Design Based on Generative Adversarial Networks," *ACS Nano*, vol. 13, no. 8, pp. 8872–8878, Aug. 2019, doi: 10.1021/acsnano.9b02371.

[64] Z. Liu, D. Zhu, S. P. Rodrigues, K.-T. Lee, and W. Cai, "Generative Model for the Inverse Design of Metasurfaces," *Nano Lett.*, vol. 18, no. 10, pp. 6570–6576, Oct. 2018, doi: 10.1021/acs.nanolett.8b03171.

[65] D. Saxena and J. Cao, "Generative Adversarial Networks (GANs): Challenges, Solutions, and Future Directions," *ACM Comput. Surv.*, vol. 54, no. 3, pp. 1–42, Apr. 2022, doi: 10.1145/3446374.

[66] J. C. Helton and F. J. Davis, "Latin hypercube sampling and the propagation of uncertainty in analyses of complex systems," *Reliability Engineering & System Safety*, vol. 81, no. 1, pp. 23–69, Jul. 2003, doi: 10.1016/S0951-8320(03)00058-9.

[67] M.-L. Chuang and M.-T. Wu, "Microstrip Diplexer Design Using Common T-Shaped Resonator," *IEEE Microw. Wireless Compon. Lett.*, vol. 21, no. 11, pp. 583–585, Nov. 2011, doi: 10.1109/LMWC.2011.2168949.

[68] Ming-Lin Chuang, "Dual-Band Impedance Transformer Using Two-Section Shunt Stubs," *IEEE Trans. Microwave Theory Techn.*, vol. 58, no. 5, pp. 1257–1263, May 2010, doi: 10.1109/TMTT.2010.2045560.

[69] K. F. Man, K. S. Tang, and S. Kwong, "Genetic algorithms: concepts and applications [in engineering design]," *IEEE Trans. Ind. Electron.*, vol. 43, no. 5, pp. 519–534, Oct. 1996, doi: 10.1109/41.538609.

[70] J. Kennedy and R. Eberhart, "Particle swarm optimization," in *Proceedings of ICNN'95 - International Conference on Neural Networks*, Perth, WA, Australia: IEEE, 1995, pp. 1942–1948. doi: 10.1109/ICNN.1995.488968.

[71] M. M. A. Ali, A. Jamali, A. Asgharnia, R. Ansari, and R. Mallipeddi, "Multi-objective Lyapunov-based controller design for nonlinear systems via genetic programming," *Neural Comput & Applic*, vol. 34, no. 2, pp. 1345–1357, Jan. 2022, doi: 10.1007/s00521-021-06453-1.

[72] D. Comaniciu and P. Meer, "Mean shift: a robust approach toward feature space analysis," *IEEE Transactions on Pattern Analysis and Machine Intelligence*, vol. 24, no. 5, pp. 603–619, 2002, doi: 10.1109/34.1000236.

[73] D. Kim and D. R. Englund, "Quantum reference beacon–guided superresolution optical focusing in complex media," *Science*, vol. 363, no. 6426, pp. 528–531, 2019, doi: 10.1126/science.aar8609.

[74] D. S. Wiersma, "The physics and applications of random lasers," *Nature Physics*, vol. 4, no. 5, pp. 359–367, May 2008, doi: 10.1038/nphys971.

[75] N. M. Estakhri, N. Mohammadi Estakhri, and T. B. Norris, "Emergence of coherent backscattering from sparse and finite disordered media," *Scientific Reports*, vol. 12, no. 1, p. 22256, Dec. 2022, doi: 10.1038/s41598-022-25465-y.

[76] N. M. Estakhri and T. B. Norris, "Coherent Two-photon Backscattering and Induced Angular Quantum Correlations in Multiple-Scattered Two-Photon States of the Light." 2024.

[77] J. Almutlaq *et al.*, "Engineering colloidal semiconductor nanocrystals for quantum information processing," *Nature Nanotechnology*, Mar. 2024, doi: 10.1038/s41565-024-01606-4.

[78] H. Nasari, M. Dupré, and B. Kanté, "Efficient design of random metasurfaces," *Opt. Lett.*, vol. 43, no. 23, pp. 5829–5832, Dec. 2018, doi: 10.1364/OL.43.005829.

# Supplementary Material

# Efficient Inverse Design of Plasmonic Patch Nanoantennas using Deep Learning

**Saeed Hemayat,**[a†] **Sina Moayed Baharlou,**[a,b†] **Alexander Sergienko,**[b] **and Abdoulaye Ndao**[a,b*]

[a] Department of Electrical and Computer Engineering, University of California San Diego, La Jolla, CA 92093, USA
[b] Department of Electrical and Computer Engineering and Photonics Center, Boston University, 8 Saint Mary's Street, Boston, MA 02215, USA
* Corresponding author, Email: <u>a1ndao@ucsd.edu</u>
[†] Equal contribution

**The supplementary file is organized as follows:**

**S1** Dataset Generation and Analysis
**S2** Basic Antenna Design
**S3** Alternative Approach to Model the MIM Structures
**S4** Obtaining Widths of the Patches
**S5** Proposed Architecture and Hyperparameter Optimization
**S6** Extended Results
**S7** Nonlinearity of the problem
**S8** Comparison to lookup table algorithm

## S1  Dataset Generation and Analysis

*1.1  Latin Hypercube Sampling*

Latin hypercube sampling (LHS) [1], [2], [3] behaves mostly similar to stratified sampling and efficiently reduces the number of runs required for reliable estimation of the behavior of a system. In each dimension, a value is randomly selected from each interval with a critical condition that once a value is selected from a specific interval, same interval cannot be chosen again (also known as sampling without replacement). The selected values from each dimension must be combined afterwards to form a unique sample point. Putting the above explanations in a mathematical formalism, for each dimension $i = 1, 2, ..., n$, the whole interval is divided into m equal subintervals $[(j − 1)/m , j/m]$ (where $j = 1, 2, ..., m$) with a length of $1/m$. Assuming dimension $i$ and subinterval $j$, a random point $x_{ij}$ is selected according to the following equation:

$$x_{ij} = \frac{j + r_{ij} - 1}{m}, \quad (S1)$$

where rij is a random number in the interval range. In order to create a single point across all dimensions, one must combine the stratified (equally probable) intervals with the random permutations to select one point from each interval. A point pk, as a unique combination across all dimensions can be generated considering $\xi_i$, a random permutation of {1, 2, ..., m} in the $i^{th}$ dimension using the following equation:

$$p_k = \left(x_{1\xi_{1(k)}}, x_{2\xi_{2(k)}}, \ldots, x_{n\xi_{n(k)}}\right), \quad (S2)$$

where $k = 1, 2, ..., m$. It can be seen that, a random permutation of the m intervals for each of dimensions will be generated independently for each dimension which means that each dimension will have its own unique ordering of intervals.

LHS guarantees that each subinterval contributes just one sample per each dimension, ensuring that samples are spread evenly over the entire space and not clustered together as they might in simple random sampling. It should be noted that, LHS takes advantage of a far efficient approach as it does not require as many samples to fill the space, compared to the random sampling. This is due to the fact that LHS inherently ensures the coverage across all dimensions, contrary to random sampling in high-dimensional spaces, as the volume increases exponentially with addition of dimensions.

*1.2 Dataset with four parameters*

To generate the dataset with four parameters, we followed the same process as we did for the dataset with three parameters. This dataset is comprised of 100,000 samples where 90% of the data has been used for the training (90,000 samples) and the remaining 10% has been used for validation and testing (5,000 samples each).

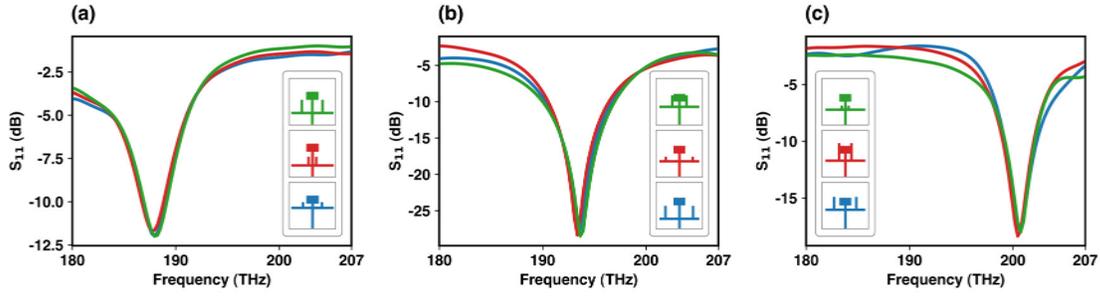

**Figure S1.** Illustration of the one-to-many mappings present in the dataset. (a-c) depicts three instances in which three distinct devices exhibit a similar $S_{11}$ response.

*1.3 One-to-many mappings*

Figure S1 illustrates three samples showing the existence of one-to-many mappings in dataset with three parameters.

*1.4 Linear Correlation*

Pearson's Product-moment correlation is used here to determine the correlation between the radiation pattern cuts at different frequencies. The radiation pattern of the samples at each frequency can be considered as a multi-dimensional random variable. Therefore, to determine the overall correlation, we have calculated Pearson's coefficient for each dimension individually and took the average across all dimensions afterward. Figure S2 displays the determined correlation matrix, where each entry shows the averaged Pearson's coefficient between two cuts at different frequencies. The results indicate a strong correlation between the radiation pattern cuts in $\varphi = 0°$ and those at $\varphi = 90°$ planes.

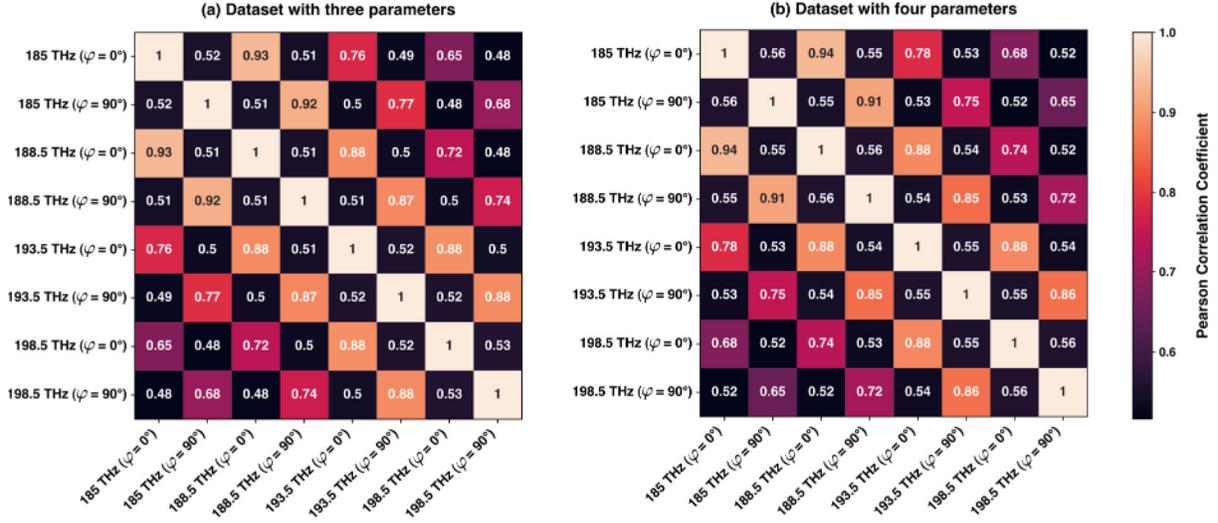

**Figure S2.** The correlation matrices showing the strong linear correlation between the radiation patterns at different frequencies in the same cut. (a) correlation matrix computed on the dataset with three parameters, (b) correlation matrix computed on the dataset with four parameters.

## S2  Basic Antenna Design

The propagating mode in a an MIM plasmonic waveguide is a TM mode, and as a result (assuming the dielectric layer is spanned over $\frac{h_d}{2} < z < \frac{h_d}{2}$ region as shown in Figure 2 of main manuscript), one can write the magnetic field as the following [4]:

$$H_x = \begin{cases} Ae^{j\beta y + \alpha_1 z}, & z < -\frac{h_d}{2} \\ Be^{j\beta y + \alpha_2 z} + B'e^{j\beta y - \alpha_2 z}, & -\frac{h_d}{2} < z < \frac{h_d}{2}, \\ A'e^{j\beta y - \alpha_1 z}, & z > +\frac{h_d}{2} \end{cases} \quad (S3)$$

where, $\beta$ is the propagation constant of the propagating plasmonic mode, $\alpha_i = \sqrt{\beta^2 - k_0^2 \varepsilon_i}$ is the decay constant in each layer, and $\varepsilon_i$ is the permittivity of the corresponding layers. The present analysis serves only to constitute an approximate relationship. Without loss of generality, we will assume the mode is symmetric which implies $A = A'$ and $B = B'$. The coefficient $B(B')$ can be determined using the boundary conditions, however we will not calculate it here since it is the shared factor in both magnetic and electric fields (as shown in the following) and will be eliminated in the calculation of the impedance. Considering the magnetic field in the insulator region and assuming a time-harmonic dependence ($e^{j\omega t}$) for both electric and magnetic fields, and using $\nabla \times \vec{H} = \frac{\partial \vec{E}}{\partial t}$ in the insulator region (where there are no free charges, hence no currents), implies $\frac{\partial H_x}{\partial y} = -j\omega\varepsilon_i E_z$ and $\frac{\partial H_x}{\partial z} = j\omega\varepsilon_i E_y$:

$$E_z = \frac{-2B\beta}{\omega\varepsilon_i} \cosh(\alpha_2 z) e^{j\beta y}, \tag{S4}$$

$$E_y = \frac{-2B\alpha_2}{j\omega\varepsilon_i} \sinh(\alpha_2 z) e^{j\beta y}. \tag{S5}$$

As mentioned earlier in this section, the propagating plasmonic mode along a MIM waveguide is transverse magnetic (TM) in nature, and as a result it cannot be described or analyzed using the conventional transmission line theory. However, one can use the following approximative approach to only obtain the basic geometrical parameters of the antenna and the feed.

In the near-infrared (NIR) regime, the absolute value of the permittivity of metals is very large compared to the permittivity of dielectrics like silicon dioxide (SiO$_2$), so that the ratio of the transverse to longitudinal component of the electric field can be written as $|E_z/E_y| = |\beta/\alpha_2||\coth(\alpha_2 z)|$, and considering $\varepsilon_m$ and $\varepsilon_d$ as the permitivitties of the metallic and dielectric layers, respectively, can be simplified to $|E_z/E_y| = |\sqrt{(\varepsilon_m/\varepsilon_d)}||\coth(\alpha_2 z)|$ which is a function that has its maximum in the middle of the insulator region. For instance, using silver ($\varepsilon_m$=-130.74 + $j$3.28 at $\lambda_0$ = 1550 nm) [5] and SiO$_2$ ($\varepsilon_d$ = 2.33 at $\lambda_0$ = 1550 nm), one can immediately realize that the ratio $\left|\sqrt{\frac{\varepsilon_m}{\varepsilon_d}}\right| \approx 7.5$ and its multiplication to the diverging (at $z$ = 0 nm) $|\coth(\alpha_2 z)|$, results in a transverse electric field component that is much larger than its longitudinal component (see

Supplementary section S3, Figure S3(a) for the present MIM geometry), where simulation results show that the transverse component of the electric field is approximately 11 times larger than the longitudinal component). Consequently, the propagating TM mode can be approximated as a TEM mode (with a small TM component), and transmission line model can be used for the basic design and determining the essential geometric parameters of the plasmonic waveguide and the patch [6], [7]. It is crucial to recognize that the derived formulas do not constitute a precise methodology for designing patch antennas in the optical regime. Their applicability is limited to specific geometries and is heavily affected by factors such as the gap between metallic plates.

Having determined $E_z$ and $H_x$ one can calculate the characteristic impedance ($Z$) of the even mode using:

$$Z = \frac{\int E_z . dz}{\oint H_x . dl}, \tag{S6}$$

where, $l$ is the closed patch encircling the conductor. Since the magnetic field has no $z$-component, hence the closed path $l$ only includes $x$-components from 0 to $W$ (width of the conductor), as a result:

$$Z = -\frac{2\beta}{\alpha_2} \frac{1}{\omega \varepsilon W} \tanh\left(\alpha_2 \frac{h_d}{2}\right). \tag{S7}$$

Having established the approximate preliminary relationships for the inductances, capacitances, and impedances, and using equation (S7), the basic optimal parameters of the patch can be chosen using this approximative method.

## S3  Alternative Approach to Model the MIM Structures

Applying an electric field to a conductive structure, enforces its electrons to accelerate, however electrons will not reach their final velocity instantaneously, as they need to accelerate gradually.

This resistance to motion due to the electron mass, combined with the fact that in plasmonic regime the inertia of the electron gas cannot be neglected (as surface plasmons are typically associated with structures that have very small feature sizes on the order of the wavelength of visible light or smaller), is the origin of kinetic inductance. This type of inductance is obvious in Drude's complex conductivity formula [8]:

$$\sigma_m(\omega) = \frac{nq^2\tau}{m(1+\omega^2\tau^2)} - j\frac{nq^2\omega\tau^2}{m(1+\omega^2\tau^2)},\tag{S8}$$

where, $n$ is the number of conduction electrons, $q$ is the charge of and electron, $\tau$ is the collision time, $m$ is the effective mass of an electron in the metal, and $\omega$ is the angular frequency. Although kinetic inductance can be neglected at frequencies less than 100 - 110 GHz since metals typically have collision times on the order of $10^{-14}$ s, multiplication of $\omega$ into the collision time leads to a very small value. However, in the optical regime where frequencies are on the orders of hundreds of THz ($\approx 10^{14}$ 1/s), $\omega\tau$ is not small anymore and the imaginary part of the conductivity cannot be neglected. For the parallel plate geometry (MIM) one must consider the kinetic inductance $L_k$, intra-plate Faraday inductance $L_{f\_i}$, and cross-plate Faraday inductance $L_{f\_c}$ [8] (it is widely known that kinetic inductance dominates over all other kind of inductances at large wavevectors, however in the NIR regime, all inductances mentioned above must be considered). The present analysis holds only when the plate spacing is not larger than the modal wavelength, otherwise, only a fraction of the electric field lines reaches from one plate to another, and the remaining field lines will only contribute to intra-plate capacitance. The $L_K$, $L_{f\_i}$, and $L_{f\_c}$ can be obtained through the following equations [8]:

$$L_k = \frac{2}{\omega^2 \delta_m W \varepsilon_0 (1-\varepsilon_m)},\tag{S9}$$

$$L_{f\_i} = \frac{\mu_0}{\beta W},\tag{S10}$$

$$L_{f_c} = \frac{\mu_0 h_d e^{\beta h_d}}{W}, \qquad (S11)$$

where, $\delta_m = \left(\beta^2 - \left(\frac{\omega^2}{c^2}\right)\right)^{-\frac{1}{2}}$ is the surface wave skin depth, $W$ is the width of the MIM structure, $\varepsilon_0$ and $\varepsilon_m$ are the vacuum and metal partitivities, respectively, $\mu_0$ is the vacuum permeability, and $h_d$ is the distance between the two metallic plates. The total inductance $L_t$, which is $L_k$ in series with the parallel equivalent of $L_{f_i}$ and $L_{f_c}$ (the current in the metallic plates must flow either in the plate or cross-plate) is:

$$L_t = \frac{1}{W}\left[\frac{\mu_0 h_d}{\beta h_d + e^{-\beta h_d}} + \frac{2}{\omega^2 \delta_m \varepsilon_0 (1-\varepsilon_m)}\right]. \qquad (S12)$$

With all of the parameters at hand and considering that the total capacitance (intra- and cross-plate capacitance) $C_t = \varepsilon_0(\beta d + e^{-kh_d})\frac{W}{h_d}$ [8], the impedance can be obtained using $Z = \sqrt{\frac{L_t}{C_t}}$.

### S4 Obtaining Widths of the Patches

As has been mentioned in the main manuscript, in the near-infrared (NIR) regime, the absolute value of the permittivity of metals is very large in comparison with the permittivity of dielectrics like silicon dioxide (SiO$_2$), so that the ratio of the transverse to longitudinal component of the electric field can be approximated as $\left|\sqrt{\frac{\varepsilon_m}{\varepsilon_d}}\right|$ [6], where $\varepsilon_m$ is the metal permittivity and $\varepsilon_d$ is permittivity of the dielectric layer sandwiched between two metallic layers. For instance, as shown in the main manuscript, using silver ($\varepsilon_m$=-130.74 + $j$3.28 at $\lambda_0$ = 1550 nm) [5] and SiO$_2$ ($\varepsilon_d$ = 2.33 at $\lambda_0$ = 1550 nm), one can immediately realize that the transverse component of the electric field is approximately 7.5 times larger than its longitudinal component. It should be noted the ratio $\left|\sqrt{\frac{\varepsilon_m}{\varepsilon_d}}\right|$ has been derived from an approximate equation of the wavevector in MIM waveguides, the ratio of the transverse to longitudinal components of the electric field in the proposed MIM waveguide is typically larger (on the order of 10 to 13). This is clearly shown in Figure S3(a), where the magnitude of the $z$- (transverse) and $y$- (longitudinal) components of the electric field are plotted versus $z$-direction. Consequently, the propagating TM mode can be approximated as a TEM mode (with a small TM component), and the transmission line model can be used for basic design and determining the essential geometric parameters of the plasmonic waveguide and the patch. However, it is crucial to recognize that this approach represents a mere approximation. The derived formulas by no means constitute a comprehensive or precise methodology for designing patch antennas in the optical regime and their applicability is constrained to specific scenarios, heavily influenced by factors such as the gap between metallic plates.

On the other hand, the effective index of the mode for the present geometry can be approximately calculated using the following equation [9]:

$$n_{eff} = \sqrt{\varepsilon_d}\left(\sqrt{1 + \frac{\lambda_0}{\pi h_d \sqrt{-\varepsilon_m}}\sqrt{1 + \frac{\varepsilon_d}{-\varepsilon_m}}}\right), \quad (S13)$$

where $\varepsilon_d$ is the permittivity of the dielectric layer, $\varepsilon_m$ is the permittivity of the metallic layers, $\lambda_0$ is the free space wavelength, and $h_d$ is the height of the dielectric layer in the MIM structure ($h_d$ = 20 nm for all of the antennas in this work). The obtained values for the effective index of the mode are plotted in Figure S3(b). Consequently, width of a patch is obtained from the following equation [10]:

$$W_p = \frac{c}{2f_r}\sqrt{\frac{2}{n_{eff}^2 + 1}}, \quad (S14)$$

where $c$ and $f_r$ are the free space velocity of light and resonant frequency of the patch, respectively.

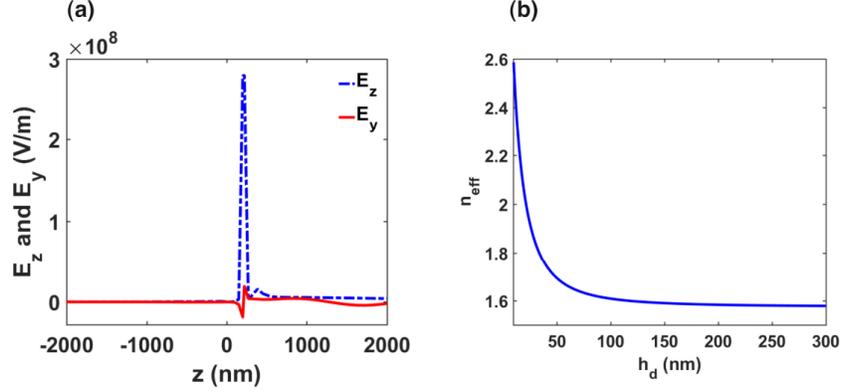

**Figure S3.** Magnitude of the transverse and longitudinal components of the electric field plotted versus a 1D cut along the $z$-axis, (b) effective index of the propagating mode inside the MIM waveguide with a width ($W_g$) = 100 nm.

## S5 Proposed Architecture and Hyperparameter Optimization

We have utilized Optuna [11], a hyperparameter optimization framework, to select the optimal architecture and hyperparameters for our neural network-based surrogate solver. Optuna provides various samplers and pruners to explore and optimize hyperparameters. It also features a web-based interface and easy-to-use visualization tools. In this work, we have utilized a pure random sampler to thoroughly explore the hyperparameter space and a median pruner to speed up the entire process.

We have performed the hyperparameter optimization for 300 trials with the following parameter distributions: the batch size has a categorical distribution with the values of (64, 128, 256, 512), learning rate has a logarithmic uniform distribution with a range from $1e^{-5}$ to $1e^{-1}$, regularization weight (only for fully-connected layers) has a logarithmic uniform distribution with a range from $1e^{-10}$ to $1e^{-6}$, the configuration of fully-connected layers has a categorical distribution with values described in Table S1, the configuration of convolutional layers has a categorical distribution with values described in Table S2, and the activation of convolutional layers has a categorical distribution with true and false values.

Table S3. shows the top five trials and their corresponding configuration and test set error. We have selected the trial (#65) for the rest of the experiments in our work.

The selected architecture has five fully connected layers, with each layer consisting of 512 neurons that use a leaky relu as the activation function. The last fully connected layer is followed by three convolutional blocks that estimate $S_{11}$ and the radiation pattern at $\varphi = 0°$ and $\varphi = 90°$.

Table S1. Configuration of fully connected layers.

| Index | Number of layers and neurons | Total number of trainable weights |
|---|---|---|
| 1 | 128×256 | 32,768 |
| 2 | 128×256×512 | 163,840 |
| 3 | 512×512 | 262,144 |
| 4 | 512×512×512 | 524,288 |
| 5 | 128×256×512×1024 | 688,128 |
| 6 | 512×512×512×512 | 786,432 |
| 7 | 512×512×512×512×512 | 1,048,576 |
| 8 | 128×256×512×1024×1024 | 1,736,704 |

Table S2. Configuration of convolutional layers.

| Index | Input dimension | Number of layers and channels | Number of trainable weights |
|---|---|---|---|
| 1 | 6×64 | 64, 32, 16, 8 | 20,376 |
| 2 | 6×64 | 64, 64, 32, 32 | 33,888 |
| 3 | 6×128 | 128, 64, 32, 16 | 81,456 |
| 4 | 6×128 | 128, 128, 64, 64 | 135,360 |

Table S3. Configuration of top five trials.

| Trial number | Batch size | Learning rate ($e^{-4}$) | Regularization weight ($e^{-8}$) | Configuration of FC layers | Configuration of Conv layers | Enable convolutional layers | $S_{11}$ test set error | Overall test set error |
|---|---|---|---|---|---|---|---|---|
| 65 | 256 | 9.27 | 0.02 | 7 | 2 | True | 0.53 | 1.29 |
| 99 | 256 | 9.82 | 0.01 | 7 | 3 | True | 0.55 | 1.31 |
| 112 | 64 | 1.16 | 0.37 | 6 | 3 | True | 0.57 | 1.44 |
| 196 | 256 | 4.60 | 3.75 | 7 | 1 | True | 0.59 | 1.45 |
| 265 | 64 | 13.63 | 1.70 | 5 | 1 | True | 0.55 | 1.36 |

Each convolutional block consists of four layers. Each layer includes an up-sampling function, followed by a convolutional and a batch normalization layer. A leaky relu is used as an activation function at the end of each layer. The kernel size used is 3×1, and the number of channels in each layer is as follows: 64, 64, 32, 32.

Optuna also calculates the importance of each hyperparameter using the fANOVA algorithm [12]. Figure S4 illustrates the importance of our hyperparameters. As the results show, having convolutional layers after the fully connected layers significantly improves the network's accuracy.

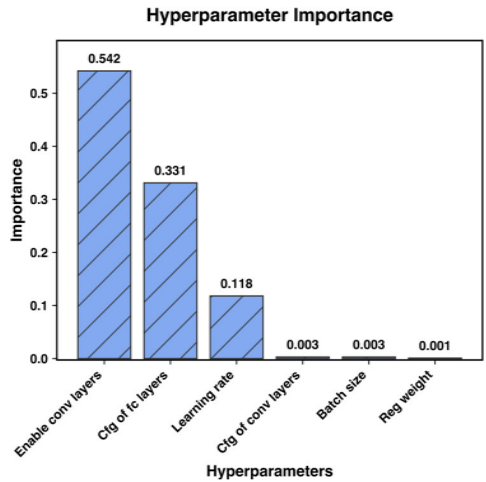

**Figure S4.** Hyperparameter importance diagram.

The number of fully-connected layers along with the learning rate are significant factors in the learning process as well. On the other hand, the configuration of convolutional layers, the batch size, and the regularization weight had no significant contribution to the overall accuracy.

**S6   Extended Results**

This section demonstrates the extended results achieved by training the proposed network on the dataset with three and four parameters. Figure S5 depicts the prediction accuracy and error distribution of the proposed surrogate solver for the device with four degrees of freedom (dataset with four parameters). The qualitative results of the inverse design verification experiment (Section 2.4.1 of main manuscript) for the device with four degrees of freedom are illustrated in Figure S6.

The error distribution of the verification experiments (Section 2.4.1 of main manuscript) for the device with three and four degrees of freedom are shown in Figure S7 and Figure S8, respectively.

Figure S9-S12 demonstrate the extended query-based results of Section 3 of the main manuscript.

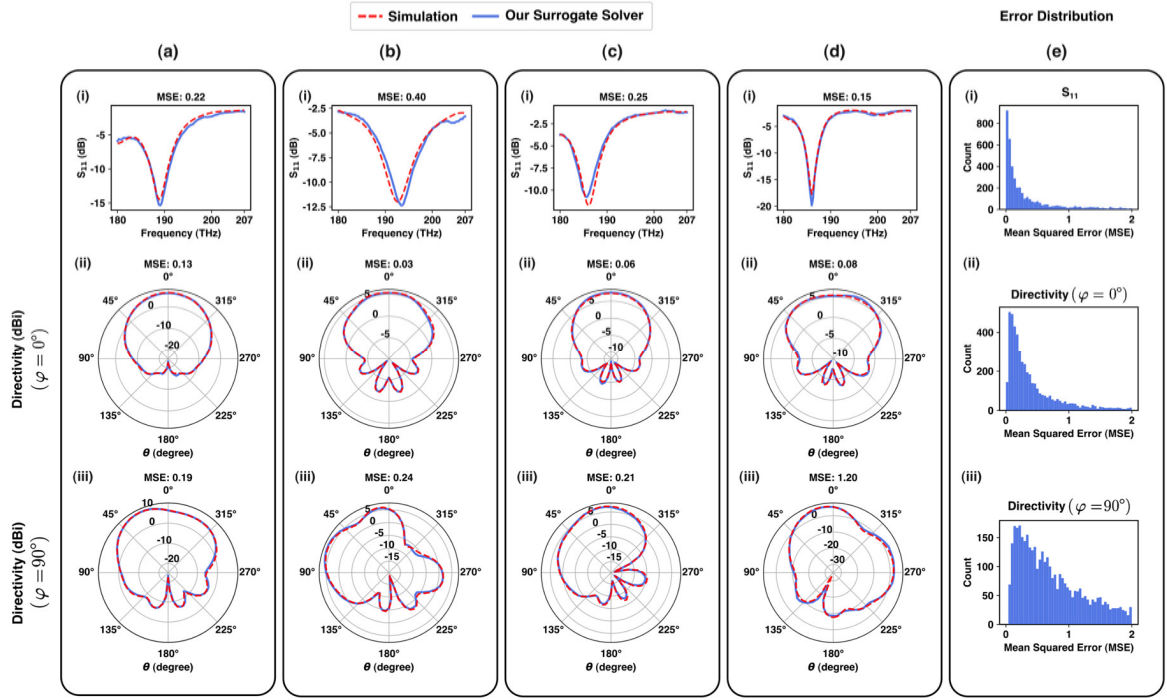

**Figure S5.** Prediction accuracy and the error distribution of the proposed surrogate solver trained on the dataset of four parameters. (a-d) the simulated response and the predicted response of four devices, (e) the error distribution of each response type.

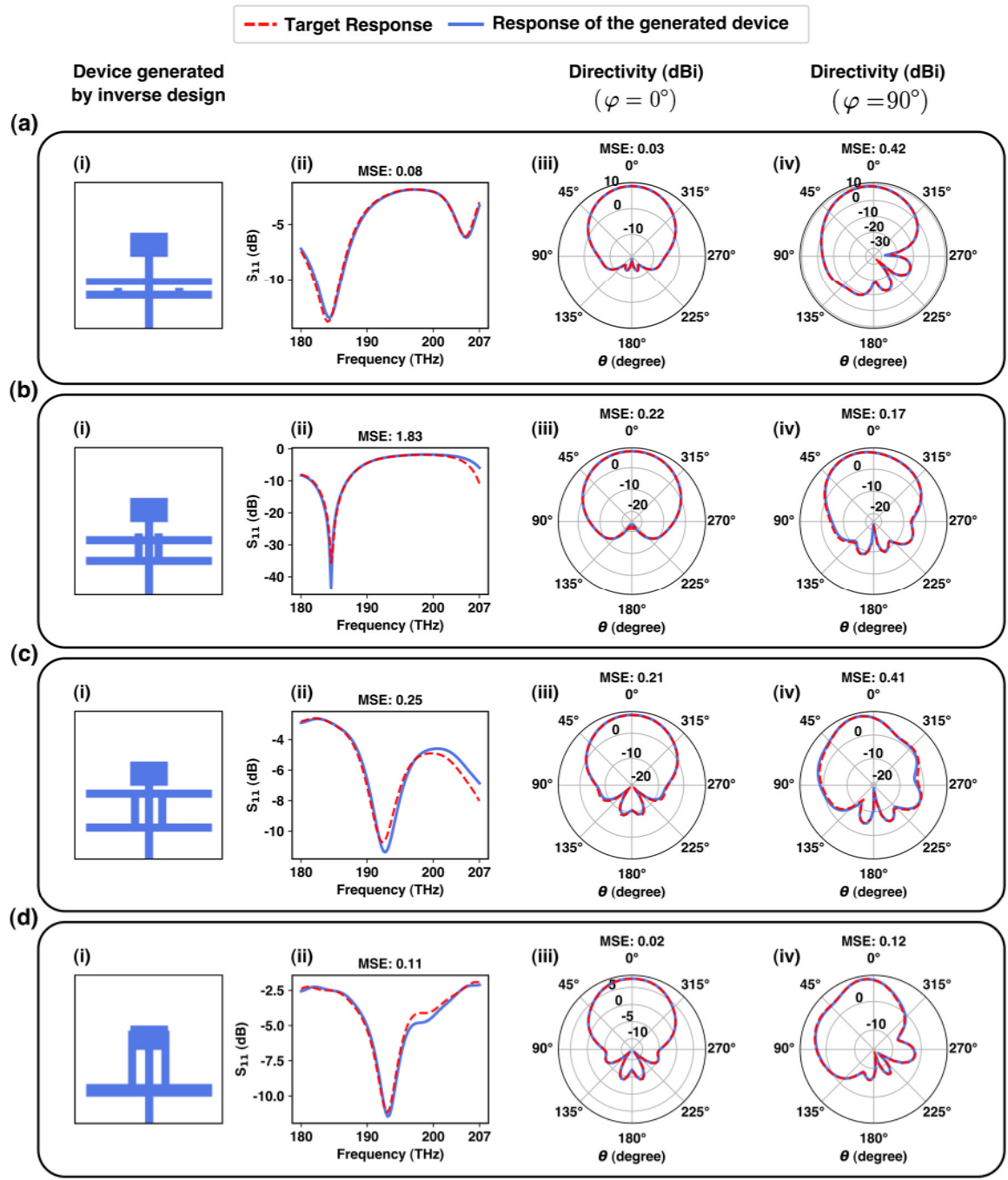

**Figure S6.** Inverse design verification experiment with the goal of generating single optimal devices given the target responses (device with four degrees of freedom).

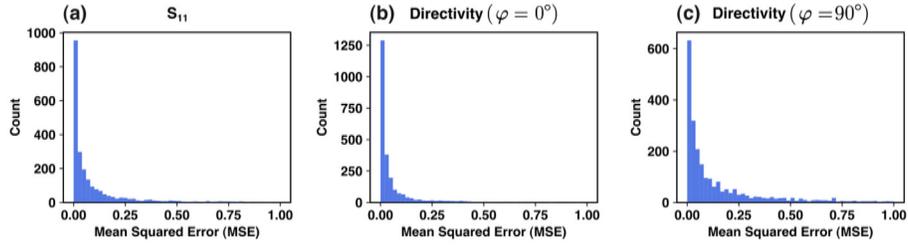

**Figure S7.** The error distribution of inverse design verification experiment for the device with three degrees of freedom.

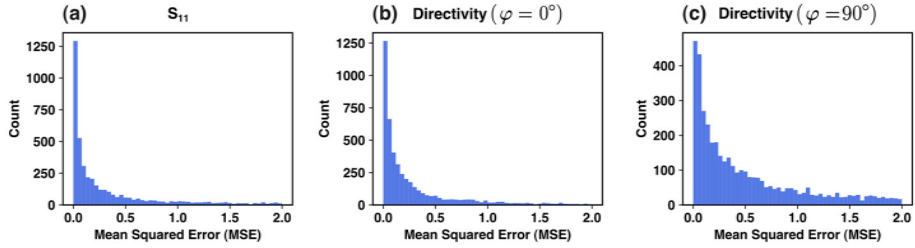

**Figure S8.** The error distribution of inverse design verification experiment for the device with four degrees of freedom.

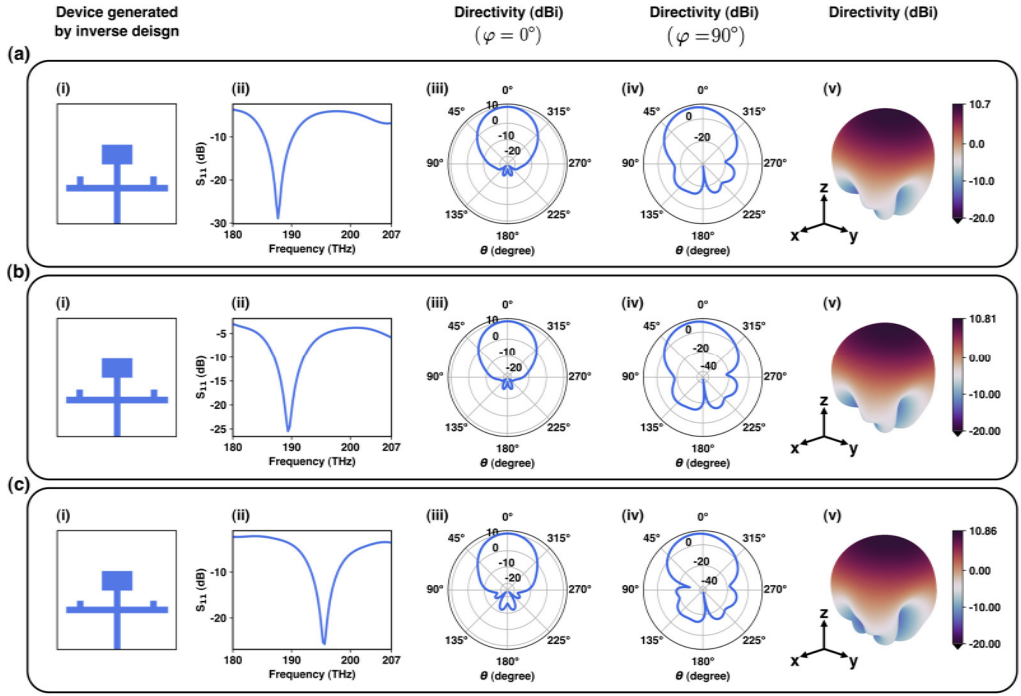

**Figure S9.** (a-c) Exemplary single band nanoantennas generated by the inverse design framework for $f$ = 188.5 THz, $f$ = 190 THz, $f$ = 195 THz with $S_{11}$ < -10 dB and the highest possible directivity in $\varphi = 0°$ and $\varphi = 90°$ planes. Each subfigure (i-v) in panels (a-c) shows the schematic of the device, $S_{11}$, directivity in $\varphi = 0°$ plane, directivity in $\varphi = 90°$ plane, and the 3D radiation pattern for each of the devices, respectively.

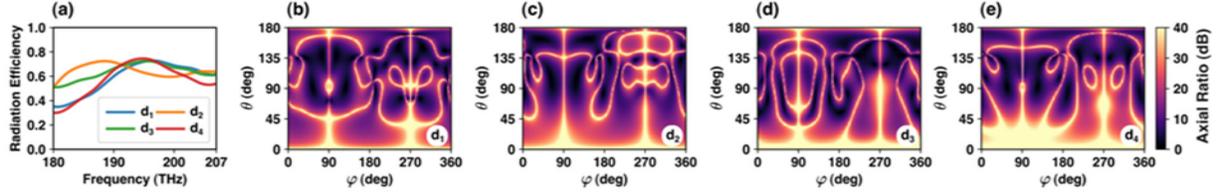

**Figure S10.** Radiation efficiency and axial ratio of the single band nanoantennas designed by the proposed inverse design framework. (a) Radiation efficiency of devices $d_1$-$d_4$ shown in Figure 7 of the main manuscript, (b-e) axial ratio of devices $d_1$-$d_4$, respectively.

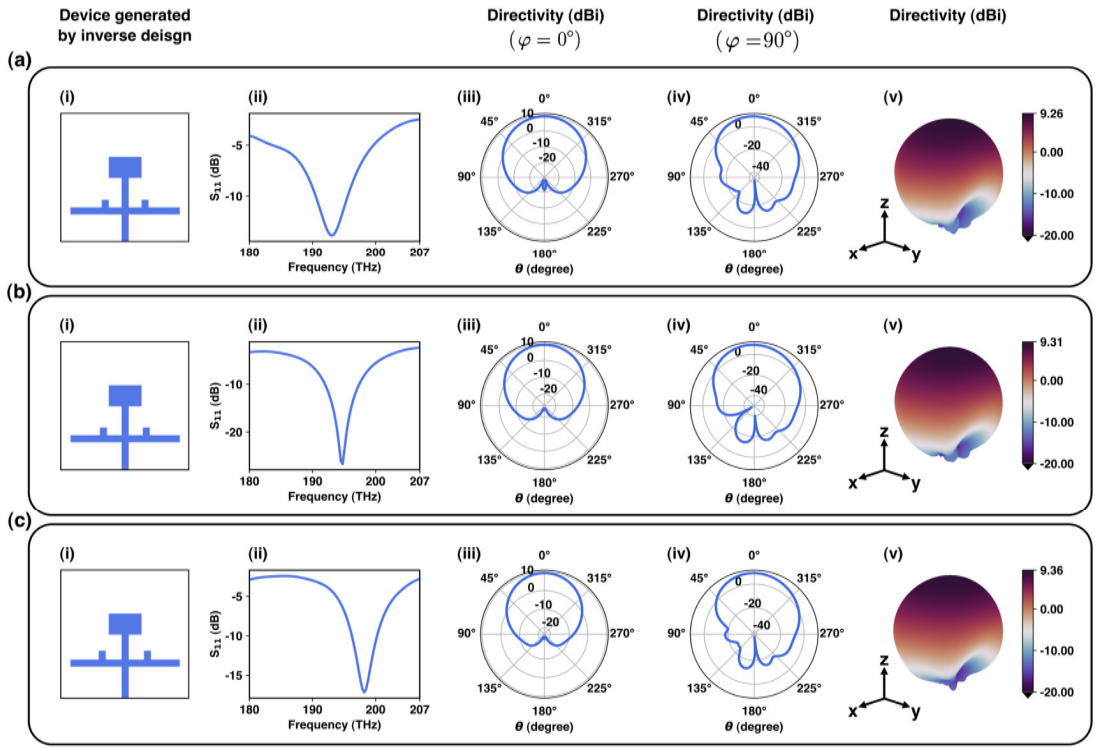

**Figure S11.** (a-c) Exemplary backlobe-suppressed single band nanoantennas generated by the inverse design framework for $f = 193.5$ THz, $f = 195$ THz, $f = 198.5$ THz with $S_{11} < -10$ dB and the highest possible directivity in $\varphi = 0°$ and $\varphi = 90°$ planes, and a suppressed radiation in $\theta = 180°$. Each subfigure (i-v) in panels (a-c) shows the schematic of the device, $S_{11}$, directivity in $\varphi = 0°$ plane, directivity in $\varphi = 90°$ plane, and the 3D radiation pattern for each of the devices, respectively.

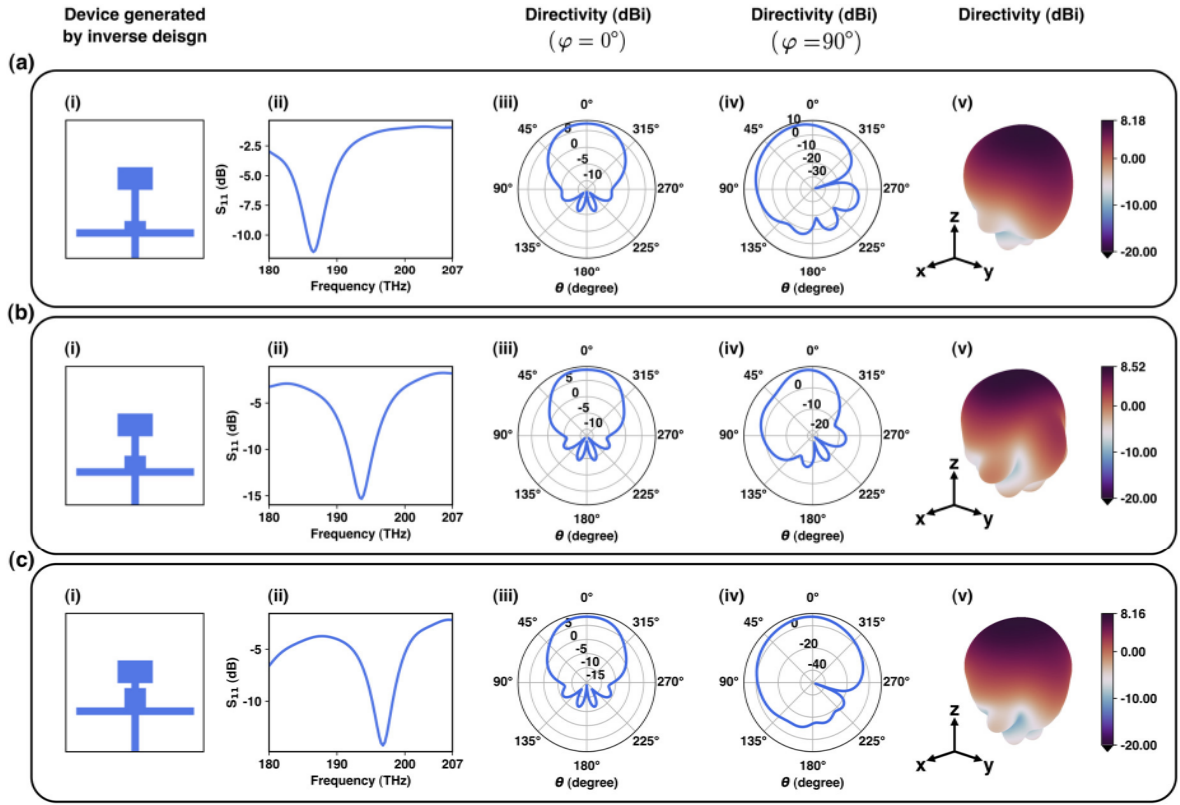

**Figure S12.** (a-c) Exemplary constraint-imposed single band nanoantennas generated by the inverse design framework, where the location of the arm of the T-stub is fixed at 50 nm and the network is tasked to generate devices with $S_{11} < $ -10 dB and directivity > 8 dBi in $\varphi = 0°$ and $\varphi = 90°$ planes, for $f$ = 186.5 THz, $f$ = 193.5 THz, $f$ = 196.5 THz. Each subfigure (i-v) in panels (a-c) shows the schematic of the device, $S_{11}$, directivity in $\varphi = 0°$ plane, directivity in $\varphi = 90°$ plane, and the 3D radiation pattern for each of the devices, respectively.

## S7 Nonlinearity of the problem

Following a through discussion on nonlinearity of the current problem despite having 3 and 4 parameters for each case, it would be beneficial to show some instances of this nonlinearity. Figure S13 shows two examples of this nonlinearity, where a slight change in the location of the arm of the T-stub ($D_a$) by steps of 20 nm, in two different nanoantennas with two sets of fixed parameters ($D_{st1}$, $L_a$) and ($D_{st1}'$, $L_a'$), leads to nonlinear changes in the $S_{11}$. The step size of 20 nm was chosen because the average distance between two closest points is about 20 nm. Assuming that there are 36.8 points in each dimension and the average range of motion of the parameters is 700 nm, this means that there is a 20 nm distance between each configuration.

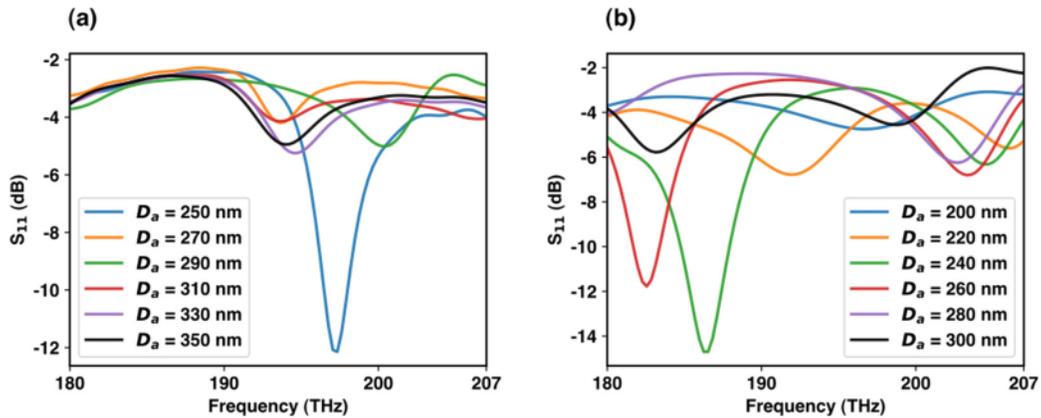

**Figure S13.** Illustration of the non-linearity of the problem, (a) and (b) $S_{11}$ of two different antennas with two sets of fixed parameters ($D_{st1}$, $L_a$) and ($D_{st1}'$, $L_a'$) for the first and the second antennas while $D_a$ is changed with steps of 20 nm, showing large variations in $S_{11}$ of each antenna with a slight change in just one parameter ($D_a$). The step size of 20 nm was chosen because the average distance between two closest points is about 20 nm. Assuming that there are 36.8 points in each dimension and the average range of motion of the parameters is 700 nm, this means that there is a 20 nm distance between each configuration.

## S8 Comparison to lookup table algorithm

The mapping between the devices and their responses is highly complex and non-linear. As a result, many devices with desirable responses are not included in the training set. However, the multi-head convolutional neural network has the capability to learn these mappings and apply them to new, unseen regions. Its performance differs from that of a simple lookup table. Figure S14 shows several examples where the surrogate solver successfully generates useful devices that were not part of the training set.

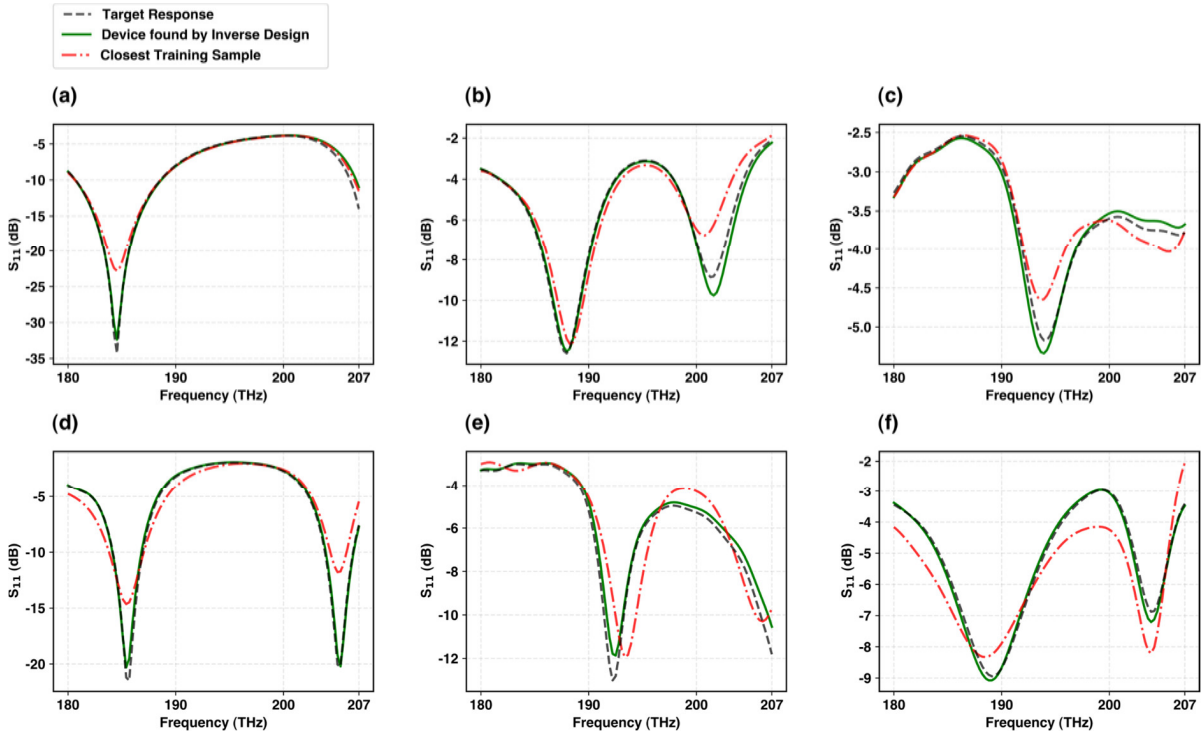

**Figure S14.** Qualitative examples showing that the surrogate solver behavior is different than a lookup algorithm. The surrogate solver has the ability to generalize to unseen data. (a-f) several instances where the surrogate solver has found a more accurate device given the target response comparing to those samples exists in the training set. The blacked dashed line shows the target response that the device must have, the green line shows the response of the device found by our inverse design framework, and the red dashed line shows the response of the closest training sample to the target response.

# References

[1] M. D. McKay, R. J. Beckman, and W. J. Conover, "A Comparison of Three Methods for Selecting Values of Input Variables in the Analysis of Output from a Computer Code," Technometrics, vol. 21, no. 2, p. 239, May 1979, doi: 10.2307/1268522.

[2] R. L. Iman, J. C. Helton, and J. E. Campbell, "An Approach to Sensitivity Analysis of Computer Models: Part I—Introduction, Input Variable Selection and Preliminary Variable Assessment," Journal of Quality Technology, vol. 13, no. 3, pp. 174–183, Jul. 1981, doi: 10.1080/00224065.1981.11978748.

[3] J. C. Helton and F. J. Davis, "Latin hypercube sampling and the propagation of uncertainty in analyses of complex systems," Reliability Engineering & System Safety, vol. 81, no. 1, pp. 23–69, Jul. 2003, doi: 10.1016/S0951-8320(03)00058-9.

[4] S. A. Maier, Plasmonics: Fundamentals and Applications. New York, NY: Springer US, 2007. doi: 10.1007/0-387-37825-1.

[5] P. B. Johnson and R. W. Christy, "Optical Constants of the Noble Metals," Phys. Rev. B, vol. 6, no. 12, pp. 4370–4379, Dec. 1972, doi: 10.1103/PhysRevB.6.4370.

[6] L. Yousefi and A. C. Foster, "Waveguide-fed optical hybrid plasmonic patch nano-antenna," Opt. Express, vol. 20, no. 16, p. 18326, Jul. 2012, doi: 10.1364/OE.20.018326.

[7] B. A. Nia, L. Yousefi, and M. Shahabadi, "Integrated Optical-Phased Array Nanoantenna System Using a Plasmonic Rotman Lens," J. Lightwave Technol., vol. 34, no. 9, pp. 2118–2126, May 2016, doi: 10.1109/JLT.2016.2520881.

[8] M. Staffaroni, J. Conway, S. Vedantam, J. Tang, and E. Yablonovitch, "Circuit analysis in metal-optics," Photonics and Nanostructures - Fundamentals and Applications, vol. 10, no. 1, pp. 166–176, Jan. 2012, doi: 10.1016/j.photonics.2011.12.002.

[9] S. Collin, F. Pardo, and J.-L. Pelouard, "Waveguiding in nanoscale metallic apertures," Opt. Express, vol. 15, no. 7, p. 4310, 2007, doi: 10.1364/OE.15.004310.

[10] C. A. Balanis, Antenna theory: analysis and design, Fourth edition. Hoboken, New Jersey: Wiley, 2016.

[11] T. Akiba, S. Sano, T. Yanase, T. Ohta, and M. Koyama, "Optuna: A Next-generation Hyperparameter Optimization Framework," in Proceedings of the 25th ACM SIGKDD International Conference on Knowledge Discovery & Data Mining, Anchorage AK USA: ACM, Jul. 2019, pp. 2623–2631. doi: 10.1145/3292500.3330701.

[12] F. Hutter, H. Hoos, and K. Leyton-Brown, "An Efficient Approach for Assessing Hyperparameter Importance," in Proceedings of the 31st International Conference on Machine Learning, E. P. Xing and T. Jebara, Eds., in Proceedings of Machine Learning Research, vol. 32. Bejing, China: PMLR, Jun. 2014, pp. 754–762. [Online]. Available: https://proceedings.mlr.press/v32/hutter14.html